%% file: main.tex
\pdfoutput=1
\documentclass[12pt,a4paper]{article}

\usepackage{ifthen} 
\newboolean{pdflatex}
\setboolean{pdflatex}{true} 

\newboolean{articletitles}
\setboolean{articletitles}{true} 

\newboolean{uprightparticles}
\setboolean{uprightparticles}{false} 

\usepackage{placeins}  
\usepackage{rotating}

\def\paperauthors{LHCb collaboration} 
\def\paperasciititle{Measurement of charm mixing and CPV with D0 to Kpipipi decays} 
\def\papertitle{Study of charm mixing and \CP violation with \Dz\to\Kpm\pimp\pipm\pimp decays} 
\def\paperkeywords{{High Energy Physics}, {LHCb}} 
\def\papercopyright{\the\year\ CERN for the benefit of the LHCb collaboration} 
\def\paperlicence{CC BY 4.0 licence}
\def\paperlicenceurl{https://creativecommons.org/licenses/by/4.0/}

\def\WSDz      {\decay{\Dz}{\Kp\pim\pip\pim}}
\def\RSDz      {\decay{\Dz}{\Km\pip\pim\pip}}
\def\bothDz      {\decay{\Dz}{\Kpm\pimp\pipm\pimp}}

\def\robsobs {{\ensuremath{ \hat{R}^{\pm}_{ij}  }}\xspace}
\def\robsprime {{\ensuremath{ R^{'\pm}_{ij}  }}\xspace}
\def\robsobs {{\ensuremath{ \hat{R}^{\pm}_{ij}  }}\xspace}
\def\robsp {{\ensuremath{ R^{+}_{ij}  }}\xspace}
\def\robsm {{\ensuremath{ R^{-}_{ij}  }}\xspace}
\def\robsobsp {{\ensuremath{ \hat{R}^{+}_{ij}  }}\xspace}
\def\robsobsm {{\ensuremath{ \hat{R}^{-}_{ij}  }}\xspace}

\def\pis {{\ensuremath{ \pi_{s} }}\xspace}

\newcommand{\aerr}[2]{{\:}^{+{\:}#1}_{-{\:}#2}}

\def\DzorDzbar {\kern \thebaroffset\optbar{\kern -\thebaroffset \PD^{0}}\xspace}

\def\rgmeas{\ensuremath{5.49\pm0.02}}
\def\kgmeas{\ensuremath{0.430\aerr{0.043}{0.039}}}
\def\dgmeas{\ensuremath{163.3\aerr{13.8}{14.8}}}

\def\xmeascp {\ensuremath{0.74\aerr{0.18}{0.25}}}
\def\ymeascp {\ensuremath{0.34\aerr{0.25}{0.29}}}
\def\xmeas   {\ensuremath{0.85\aerr{0.15}{0.24}}}
\def\ymeas   {\ensuremath{0.21\aerr{0.29}{0.27}}}
\def\dxmeas  {\ensuremath{-0.02\pm {0.04}}}
\def\dymeas  {\ensuremath{0.02\aerr{0.04}{0.03}}}

\input{preamble}
\usepackage{longtable} 
\usepackage{comment}
\usepackage{multirow}
\usepackage{booktabs}
\begin{document}

\renewcommand{\thefootnote}{\fnsymbol{footnote}}
\setcounter{footnote}{1}

\input{title-LHCb-PAPER}

\renewcommand{\thefootnote}{\arabic{footnote}}
\setcounter{footnote}{0}

\cleardoublepage

\pagestyle{plain} 
\setcounter{page}{1}
\pagenumbering{arabic}


\input{body}

\input{acknowledgements}

\FloatBarrier

\input{appendix}

\FloatBarrier

\addcontentsline{toc}{section}{References}
\bibliographystyle{LHCb}
\bibliography{main,standard,LHCb-PAPER,LHCb-CONF,LHCb-DP,LHCb-TDR}
 
\newpage
\input{Authorship_LHCb-PAPER-2025-029}

\end{document}

%% file: preamble.tex
\usepackage[top=1in, bottom=1.25in, left=1in, right=1in]{geometry}

\usepackage{microtype}
\usepackage{lineno}  
\usepackage{xspace} 
\usepackage{caption} 

\usepackage{graphicx}  
\usepackage{color}
\usepackage{colortbl}
\graphicspath{{./figs/}} 

\usepackage{amsmath} 
\usepackage{amssymb}
\usepackage{amsfonts}
\usepackage{upgreek} 

\newcommand*\patchAmsMathEnvironmentForLineno[1]{%
\expandafter\let\csname old#1\expandafter\endcsname\csname #1\endcsname
\expandafter\let\csname oldend#1\expandafter\endcsname\csname
end#1\endcsname
 \renewenvironment{#1}%
   {\linenomath\csname old#1\endcsname}%
   {\csname oldend#1\endcsname\endlinenomath}%
}
\newcommand*\patchBothAmsMathEnvironmentsForLineno[1]{%
  \patchAmsMathEnvironmentForLineno{#1}%
  \patchAmsMathEnvironmentForLineno{#1*}%
}
\AtBeginDocument{%
\patchBothAmsMathEnvironmentsForLineno{equation}%
\patchBothAmsMathEnvironmentsForLineno{align}%
\patchBothAmsMathEnvironmentsForLineno{flalign}%
\patchBothAmsMathEnvironmentsForLineno{alignat}%
\patchBothAmsMathEnvironmentsForLineno{gather}%
\patchBothAmsMathEnvironmentsForLineno{multline}%
\patchBothAmsMathEnvironmentsForLineno{eqnarray}%
}

\usepackage[pdftex,
            pdfauthor={\paperauthors},
            pdftitle={\paperasciititle},
            pdfkeywords={\paperkeywords}]{hyperref}
\usepackage{hyperxmp}
\hypersetup{
    pdfcopyright={Copyright (C) \papercopyright},
    pdflicenseurl={\paperlicenceurl}
}

\usepackage[colorinlistoftodos,textsize=scriptsize]{todonotes}

\usepackage[bottom,flushmargin,hang,multiple]{footmisc}

\usepackage[all]{hypcap} 

\input{lhcb-symbols-def} 

\hypersetup{
  colorlinks   = true, 
  urlcolor     = blue, 
  linkcolor    = blue, 
  citecolor    = red   
}

\usepackage{cite} 
\usepackage{mciteplus}

%% file: lhcb-symbols-def.tex
\usepackage{xspace} 
\usepackage{upgreek}

\def\lhcb   {\mbox{LHCb}\xspace}

\def\MagUp {\mbox{\em Mag\kern -0.05em Up}\xspace}

\ifthenelse{\boolean{uprightparticles}}%
{

 \def\Ppi         {\ensuremath{\uppi}\xspace}

 \def\PDelta      {\ensuremath{\Delta}\xspace}                 
 \def\PXi         {\ensuremath{\Xi}\xspace}                 
 \def\PLambda     {\ensuremath{\Lambda}\xspace}                 
 \def\PSigma      {\ensuremath{\Sigma}\xspace}                 
 \def\POmega      {\ensuremath{\Omega}\xspace}                 
 \def\PUpsilon    {\ensuremath{\Upsilon}\xspace}

 \def\PB      {\ensuremath{\mathrm{B}}\xspace}                 
                  
 \def\PD      {\ensuremath{\mathrm{D}}\xspace}

 \def\PK      {\ensuremath{\mathrm{K}}\xspace}

 \def\Pb      {\ensuremath{\mathrm{b}}\xspace}                 
 \def\Pc      {\ensuremath{\mathrm{c}}\xspace}

 \def\Pi      {\ensuremath{\mathrm{i}}\xspace}

 \def\Ps      {\ensuremath{\mathrm{s}}\xspace}

 \def\thebaroffset{0.0em}
}
{

 \def\Ppi         {\ensuremath{\pi}\xspace}

 \mathchardef\PDelta="7101
 \mathchardef\PXi="7104
 \mathchardef\PLambda="7103
 \mathchardef\PSigma="7106
 \mathchardef\POmega="710A
 \mathchardef\PUpsilon="7107
                  
 \def\PB      {\ensuremath{B}\xspace}                 
                  
 \def\PD      {\ensuremath{D}\xspace}

 \def\PK      {\ensuremath{K}\xspace}

 \def\Pb      {\ensuremath{b}\xspace}                 
 \def\Pc      {\ensuremath{c}\xspace}

 \def\Pi      {\ensuremath{i}\xspace}

 \def\Ps      {\ensuremath{s}\xspace}

 \def\thebaroffset{0.18em}
}
\newcommand{\offsetoverline}[2][\thebaroffset]{\kern #1\overline{\kern -#1 #2}}%

\makeatletter
\ifcase \@ptsize \relax
  \newcommand{\miniscule}{\@setfontsize\miniscule{4}{5}}
\or
  \newcommand{\miniscule}{\@setfontsize\miniscule{5}{6}}
\or
  \newcommand{\miniscule}{\@setfontsize\miniscule{5}{6}}
\fi
\makeatother

\DeclareRobustCommand{\optbar}[1]{\shortstack{{\miniscule (\rule[.5ex]{1.25em}{.18mm})}
  \\ [-.7ex] $#1$}}

\def\squark    {{\ensuremath{\Ps}}\xspace}

\def\cquark    {{\ensuremath{\Pc}}\xspace}

\def\bquark    {{\ensuremath{\Pb}}\xspace}

\def\pion   {{\ensuremath{\Ppi}}\xspace}
\def\piz    {{\ensuremath{\pion^0}}\xspace}
\def\pip    {{\ensuremath{\pion^+}}\xspace}
\def\pim    {{\ensuremath{\pion^-}}\xspace}
\def\pipm   {{\ensuremath{\pion^\pm}}\xspace}
\def\pimp   {{\ensuremath{\pion^\mp}}\xspace}

\def\kaon    {{\ensuremath{\PK}}\xspace}

\def\KorKbar {\kern \thebaroffset\optbar{\kern -\thebaroffset \PK}{}\xspace}

\def\Kp      {{\ensuremath{\kaon^+}}\xspace}
\def\Km      {{\ensuremath{\kaon^-}}\xspace}
\def\Kpm     {{\ensuremath{\kaon^\pm}}\xspace}

\def\KS      {{\ensuremath{\kaon^0_{\mathrm{S}}}}\xspace}

\def\Dbar    {{\ensuremath{\offsetoverline{\PD}}}\xspace}
\def\D       {{\ensuremath{\PD}}\xspace}

\def\DorDbar {\kern \thebaroffset\optbar{\kern -\thebaroffset \PD}\xspace}
\def\Dz      {{\ensuremath{\D^0}}\xspace}
\def\Dzb     {{\ensuremath{\Dbar{}^0}}\xspace}
\def\Dp      {{\ensuremath{\D^+}}\xspace}
\def\Dm      {{\ensuremath{\D^-}}\xspace}

\def\DpDm    {\ensuremath{\Dp {\kern -0.16em \Dm}}\xspace}

\def\Dstarp  {{\ensuremath{\D^{*+}}}\xspace}

\def\Dstarpm {{\ensuremath{\D^{*\pm}}}\xspace}

\def\theDstarp{{\ensuremath{\D^{*}(2010)^{+}}}\xspace}

\def\B       {{\ensuremath{\PB}}\xspace}

\def\BorBbar {\kern \thebaroffset\optbar{\kern -\thebaroffset \PB}\xspace}

\def\Bd      {{\ensuremath{\B^0}}\xspace}

\def\BdorBdbar {\kern \thebaroffset\optbar{\kern -\thebaroffset \Bd}\xspace}

\def\Bs      {{\ensuremath{\B^0_\squark}}\xspace}

\def\BsorBsbar {\kern \thebaroffset\optbar{\kern -\thebaroffset \Bs}\xspace}

\def\Y#1S{\ensuremath{\PUpsilon{(#1S)}}\xspace}

\def\LorLbar     {\kern \thebaroffset\optbar{\kern -\thebaroffset \PLambda}\xspace}

\newcommand{\decay}[2]{\ensuremath{#1\!\to #2}\xspace} 

\def\to                 {\ensuremath{\rightarrow}\xspace}

\def\CP                {{\ensuremath{C\!P}}\xspace}

\newcommand{\dm}{{\ensuremath{\Delta m}}\xspace}

\def\AT#1     {\ensuremath{A_{\mathrm{T}}^{#1}}\xspace}           

\def\C#1      {\ensuremath{\mathcal{C}_{#1}}\xspace}                       
\def\Cp#1     {\ensuremath{\mathcal{C}_{#1}^{'}}\xspace}                    
\def\Ceff#1   {\ensuremath{\mathcal{C}_{#1}^{\mathrm{(eff)}}}\xspace}        
\def\Cpeff#1  {\ensuremath{\mathcal{C}_{#1}^{'\mathrm{(eff)}}}\xspace}       
\def\Ope#1    {\ensuremath{\mathcal{O}_{#1}}\xspace}                       
\def\Opep#1   {\ensuremath{\mathcal{O}_{#1}^{'}}\xspace}                    

\newcommand{\ket}[1]{\ensuremath{|#1\rangle}}              

\newcommand{\nospaceunit}[1]{\ensuremath{\text{#1}}}       
\newcommand{\aunit}[1]{\ensuremath{\text{\,#1}}}       

\newcommand{\tev}{\aunit{Te\kern -0.1em V}\xspace}
\newcommand{\gev}{\aunit{Ge\kern -0.1em V}\xspace}
\newcommand{\mev}{\aunit{Me\kern -0.1em V}\xspace}
\newcommand{\kev}{\aunit{ke\kern -0.1em V}\xspace}
\newcommand{\ev}{\aunit{e\kern -0.1em V}\xspace}
 
\newcommand{\mevc}{\ensuremath{\aunit{Me\kern -0.1em V\!/}c}\xspace}
\newcommand{\gevc}{\ensuremath{\aunit{Ge\kern -0.1em V\!/}c}\xspace}
\newcommand{\mevcc}{\ensuremath{\aunit{Me\kern -0.1em V\!/}c^2}\xspace}
\newcommand{\gevcc}{\ensuremath{\aunit{Ge\kern -0.1em V\!/}c^2}\xspace}

\def\mum  {\ensuremath{\,\upmu\nospaceunit{m}}\xspace}

\def\fb   {\ensuremath{\aunit{fb}}\xspace}
\def\invfb   {\ensuremath{\fb^{-1}}\xspace}

\newcommand{\chisq}{\ensuremath{\chi^2}\xspace}

\newcommand{\chisqip}{\ensuremath{\chi^2_{\text{IP}}}\xspace}

\def\gsim{{~\raise.15em\hbox{$>$}\kern-.85em
          \lower.35em\hbox{$\sim$}~}\xspace}
\def\lsim{{~\raise.15em\hbox{$<$}\kern-.85em
          \lower.35em\hbox{$\sim$}~}\xspace}

\def\pt         {\ensuremath{p_{\mathrm{T}}}\xspace}

\def\ptot       {\ensuremath{p}\xspace}

\def\evtgen     {\mbox{\textsc{EvtGen}}\xspace}

\def\geant      {\mbox{\textsc{Geant4}}\xspace}

\def\pythia     {\mbox{\textsc{Pythia}}\xspace}

\def\tell1  {TELL1\xspace}
\def\ukl1   {UKL1\xspace}

\newcommand{\vs}{\mbox{\itshape vs.}\xspace}

\newcommand{\lhcborcid}[1]{\href{https://orcid.org/#1}{\hspace*{0.1em}\raisebox{-0.45ex}{\includegraphics[width=1em]{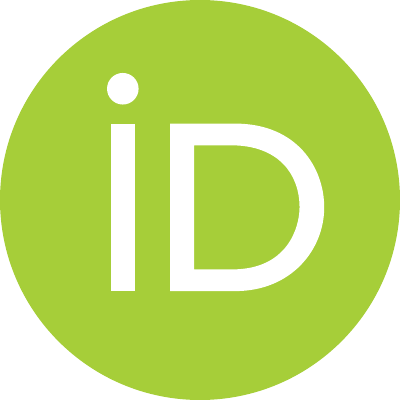}}}}

%% file: title-LHCb-PAPER.tex

\begin{titlepage}
\pagenumbering{roman}

\vspace*{-1.5cm}
\centerline{\large EUROPEAN ORGANIZATION FOR NUCLEAR RESEARCH (CERN)}
\vspace*{1.5cm}
\noindent
\begin{tabular*}{\linewidth}{lc@{\extracolsep{\fill}}r@{\extracolsep{0pt}}}
\ifthenelse{\boolean{pdflatex}}
{\vspace*{-1.5cm}\mbox{\!\!\!\includegraphics[width=.14\textwidth]{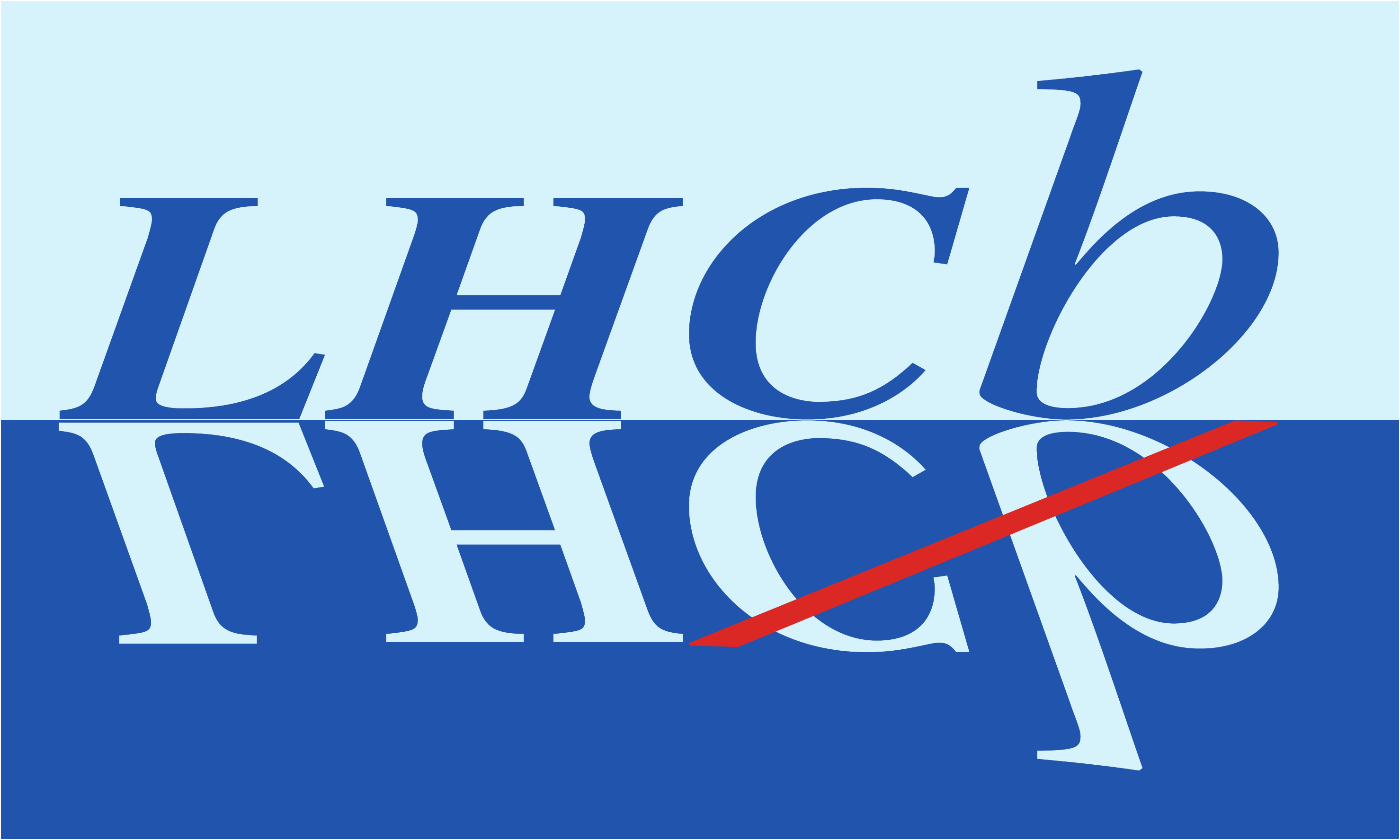}} & &}%
{\vspace*{-1.2cm}\mbox{\!\!\!\includegraphics[width=.12\textwidth]{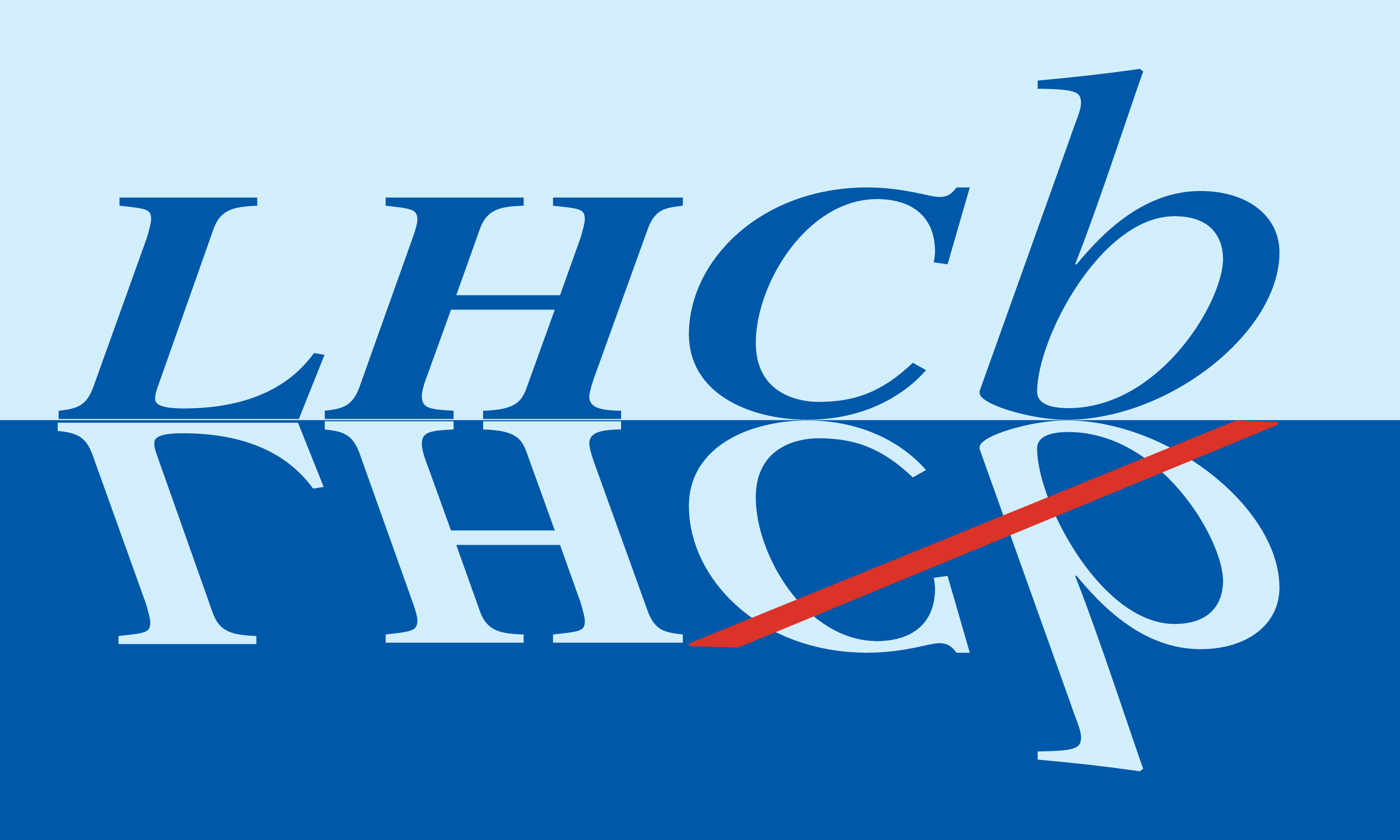}} & &}%
\\
 & & CERN-EP-2025-220 \\  
 & & LHCb-PAPER-2025-029 \\  
 & & \today \\ 
 & & \\
\end{tabular*}

\vspace*{2.0cm}

{\normalfont\bfseries\boldmath\huge
\begin{center}
  \papertitle 
\end{center}
}

\vspace*{2.0cm}

\begin{center}
\paperauthors\footnote{Authors are listed at the end of this paper.}
\end{center}

\vspace{\fill}

\begin{abstract}
  \noindent
A study of charm mixing and \CP violation in \bothDz decays is performed using data collected by the LHCb experiment in proton-proton collisions from 2015 to 2018, corresponding to an integrated luminosity of 6\invfb. The ratio of promptly produced \WSDz to \RSDz decay rates is measured as a function of \Dz decay time, both inclusive over phase space and in bins of phase space.
Taking external inputs for the \Dz--\Dzb mixing parameters $x$ and $y$ allows constraints to be obtained on the hadronic parameters of the charm decay.   When combined with previous measurements from charm-threshold experiments and at LHCb, improved knowledge is obtained for these parameters, which is valuable for studies of the angle $\gamma$ of the Unitarity Triangle.
An alternative analysis is also performed, in which external inputs are taken for the hadronic parameters, and the mixing parameters are determined, including $\Delta x$ and $\Delta y$, which are nonzero in the presence of \CP violation.  It is found that $x=\left(\xmeas\right)\%$, $y=\left(\ymeas\right)\%$, $\Delta x=\left(\dxmeas\right)\% $ and $\Delta y=\left(\dymeas\right)\%$. These results are consistent with previous measurements and the hypothesis of \CP conservation.

\end{abstract}

\vspace*{2.0cm}

\begin{center}
  Published in JHEP 12 (2025) 153
\end{center}

\vspace{\fill}

{\footnotesize 
\centerline{\copyright~\papercopyright. \href{\paperlicenceurl}{\paperlicence}.}}
\vspace*{2mm}

\end{titlepage}


\newpage
\setcounter{page}{2}
\mbox{~}
%
%
%
%

%% file: body.tex
\section{Introduction}
\label{sec:Introduction}

Neutral charm mesons can oscillate into their own antiparticles, as the mass eigenstates are linear combinations of the flavour eigenstates.  The Standard Model (SM) does not permit flavour-changing neutral currents at leading order, and thus studying these loop-level mixing processes provides sensitivity to the possible virtual contribution of currently unknown high-mass particles~\cite{Golowich:2007ka,Isidori:2010kg,UTfit:2007eik}.

In the limit of \CP symmetry, charm oscillations are characterised by the dimensionless parameters $x \equiv (m_1 - m_2)/\Gamma$ and $y \equiv (\Gamma_1 - \Gamma_2)/2\Gamma$, where $m_{1(2)}$ and $\Gamma_{1(2)}$ are the mass and decay width of the mass eigenstate $D_{1(2)}$, respectively, and $\Gamma \equiv (\Gamma_1 + \Gamma_2)/2$.  The mass eigenstate $D_{1(2)}$ here is identical to the \CP-even (odd) eigenstate.  
The current ensemble of charm-mixing and related measurements from the LHCb collaboration yields $x=(0.41 \pm 0.05)\%$ and $y=(0.621\aerr{0.022}{0.021})\%$~\cite{LHCb-CONF-2024-004,LHCb-PAPER-2021-033}.

The phenomenon of \CP violation has been observed in the decay of charm mesons in processes mediated by singly Cabibbo-suppressed amplitudes~\cite{LHCb-PAPER-2019-006}, but is expected to be negligible in decays that involve Cabibbo-favoured (CF) or doubly Cabibbo-suppressed (DCS) amplitudes~\cite{Bergmann:1999pm}.
In the case where a neutral charm meson \D is reconstructed in a final state that is accessible through both \Dz and \Dzb decays, \CP violation may, however, occur in either the mixing process or in the interference between the mixing and decay.  These categories of \CP violation have not been observed and are expected to be very small, but could be enhanced by physics beyond the SM~\cite{Kagan:2020vri,Lenz:2020awd}.  The presence of \CP violation in mixing or in the interference between mixing and decay would manifest itself through nonzero values of $\Delta x \equiv \frac{1}{2}(x^+ - x^-)$ or $\Delta y \equiv \frac{1}{2}(y^+ - y^-)$, where the $\pm$ superscripts indicate a measurement of the mixing parameters performed in separate $\CP$-conjugated processes.

A powerful method to probe for mixing and \CP violation in the charm system is to perform a study of pairs of CF and DCS decays in which interference between the DCS amplitude and the product of the CF and mixing amplitudes gives rise to effects that vary with decay time of the meson.  The decays $\Dz \to K^\pm \pi^\mp$ have been exploited by LHCb and other experiments for this purpose~\cite{LHCb-PAPER-2012-038,LHCb-PAPER-2016-033,LHCb-PAPER-2017-046,LHCb-PAPER-2024-008,LHCb-PAPER-2024-044,BaBar:2007kib,CDF:2013gvz,Belle:2014yoi}.
Inclusion of the charge-conjugate process is implied unless stated otherwise.
Any multibody decay, such as \bothDz, can be harnessed in an analogous manner, but here the four-body final state gives rise to interference effects that also vary over phase space.
The transition \RSDz is known as the `right-sign' (RS) decay and is dominant, being wholly mediated by CF amplitudes in the limit of no mixing, whereas the suppressed `wrong-sign' (WS) decay \WSDz occurs mostly through DCS amplitudes.  The presence of mixing means that $R(t)$,  the ratio of the two decay rates,  evolves with decay time $t$ and, in the limit of \CP conservation, is given by  
\begin{align}
R(t) & \equiv \frac{\Gamma[\WSDz](t)}{\Gamma [\RSDz](t)} \approx r^2 - r \kappa y' \frac{t}{\tau} + \frac{x^2 + y^2}{4} \left( \frac{t}{\tau} \right)^2, 
\label{eqn:mixing}
\end{align}
where $r$ is the magnitude of the ratio of the DCS to  CF amplitudes, averaged over the phase space of the final state, and $\tau$ is the \Dz lifetime~\cite{Harnew:2013wea,Harnew:2014zla}.
The factor $y'$ relates the mixing parameters to the phase difference $\delta$ between the CF and DCS decay amplitudes averaged over phase space and is given by $y' \equiv y \cos \delta - x \sin \delta$.\footnote{Here the convention $\CP \ket{\Dz} = +\ket{\Dzb}$ is adopted, which determines the sign of the linear term in Eq.~\ref{eqn:mixing}.} 
The coherence factor, $\kappa$, is defined by $\kappa e^{i\delta} \equiv \langle \cos \delta_l \rangle + i \langle \sin \delta_l \rangle$, where $\delta_l$ is the phase difference at a local point $l$ in phase space, and $\langle \cos \delta_l \rangle$ and $ \langle \sin \delta_l \rangle$ are the average values, over phase space, of the cosine and sine of this local phase difference~\cite{Atwood:2003mj}. This parameter is a real number between 0 and 1, which is specific to the final state, and represents a dilution of the quantum interference due to the resonant substructure of the \bothDz decays.  The coherence factor and average phase difference are of additional interest, as they enter into the expressions that govern the determination of the \CP-violating phase $\gamma$ in $B^- \to DK^-$ decays, where the neutral charm meson $D$ is reconstructed in the final state $K^\pm \pi^\mp \pip \pim$~\cite{Atwood:2003mj}. 

This paper presents a study of charm mixing and \CP violation in the decays {\mbox \bothDz}. The analysis is based on data collected by the LHCb experiment in proton-proton ($pp$) collisions at $\sqrt{s} = 13\tev$ during Run~2 of the LHC (2015--2018), corresponding to an integrated luminosity of~6\invfb. The \Dz mesons arise from the strong decays $\theDstarp \to \Dz \pip$, where the \Dstarp mesons are produced promptly in the $pp$ collision.  Analyses are performed for both the inclusive final state and in intervals~(bins) of phase space.  The measured observables are interpreted in terms of  the coherence factor and average phase of the charm-meson decay, and 
also in terms of the charm-mixing and \CP-violation parameters.
A previous analysis of the same channel was reported using data collected from Run~1~\cite{LHCb-PAPER-2015-057}, during 2010--2012, but this study was restricted to the inclusive final state and did not measure any observables sensitive to \CP violation.

\section{Measurement strategy}

Any measurement involving \bothDz decays that is inclusive over phase space dilutes interference effects, due to the variation in the phase difference between \Dz and \Dzb decay amplitudes.  It was therefore proposed in Ref.~\cite{Evans2020_new} to partition the phase space into four bins, with the bin boundaries chosen to minimise the dilution of the interference effects, and thereby provide high sensitivity to the angle $\gamma$ of the Unitarity Triangle in an analysis of $B^- \to DK^-$, $D \to K^\pm \pi^\mp \pip \pim$ decays.  
The binning scheme is constructed according to the prediction of amplitude models of the resonant substructure of the charm-meson decays developed by the LHCb collaboration~\cite{LHCb-PAPER-2017-040}, and has been exploited in a precise measurement of the angle $\gamma$ with LHCb data~\cite{LHCb-PAPER-2022-017}. 

The same binning scheme is used to perform the charm-mixing measurements presented in this paper.  As well as binning the data in phase space, a binning is also made in decay time.  Candidates are partitioned into ten bins of decay time in the range $0.24 < t/\tau < 8.00$, with the boundaries chosen to ensure approximately equal numbers of candidates in each bin.  Fewer than one in 10\,000 candidates are excluded by the lower bound on the range, as the trigger and selection requirements described in Sec.~\ref{sec:selection} already remove decays at low times. The upper bound suppresses candidates from decays of $b$~hadrons.  The measurement then proceeds by determining the ratio of WS to RS decay rates in bin $i$ of phase space and $j$ of decay time, with Eq.~\ref{eqn:mixing} becoming 
\begin{align}
R_{ij} & \equiv \frac{\Gamma[\WSDz]_{ij}}{\Gamma [\RSDz]_{ij}} \approx (r_i)^2 - r_i (\kappa y')_i \left< \frac{t_{ij}}{\tau} \right> + \frac{x^2 + y^2}{4} \left< \frac{t_{ij}^2}{\tau^2} \right>,
\label{eqn:mixing_binned}
\end{align}
where $\left<t_{ij}^{(2)}/\tau^{(2)}\right>$ is the average normalised decay time (squared) of the candidates in bins $i,j$, as measured in the RS sample.
It is noted that the amplitude ratio $r_i$ and product of parameters $(\kappa y')_i$ are designated with a subscript indicating their phase-space bin, as these quantities are expected to vary with phase space.  A fit to the set of measurements of $R_{ij}$ across bins of decay time accesses the observables $r_i$ and $(\kappa y')_i$ for phase-space bin $i$, together with $x^2 + y^2$.  From these observables, $x$ and $y$ may be determined, by taking as input the values of the coherence factors and average phase differences obtained from studies of quantum-correlated $\D \Dbar$ decays at charm threshold, performed with data accumulated by the BESIII and CLEO-c experiments~\cite{BESIII2021_new,Evans2020_new}. The current knowledge of these parameters from charm-threshold studies, and including phase-space inclusive constraints from an earlier LHCb analysis~\cite{LHCb-PAPER-2015-057}, is summarised in Table~\ref{tab:coherence_factors}. Conversely, it is possible to derive constraints on the coherence factors and average phase differences, by using external measurements of the mixing parameters as input~\cite{LHCb-CONF-2024-004}. This latter approach has the potential to improve on the knowledge obtained from the charm-threshold experiments alone, and thereby benefit the measurement of the angle $\gamma$. All charm mixing and hadronic parameters are determined by minimising a global $\chi^2$, composed of contributions from the charm mixing and threshold datasets, which is discussed in Sect.~\ref{sec:interpretation}.

\input{bin_definition}

An inclusive analysis is also performed, as it remains desirable to perform a determination of the charm-mixing observables integrated over phase space. Although such an analysis has lower sensitivity to the mixing parameters, 
the constraints on the coherence factor and average phase difference 
are valuable input for many measurements, in particular those where limited sample size does not allow for the phase space to be binned.  
The current knowledge of these inclusive parameters from the charm-threshold experiments and the earlier LHCb analysis is also included in Table~\ref{tab:coherence_factors}.   
Note that the sum over the four bins does not encompass all phase space, as it excludes a small region in which at least one of the $\pip\pim$ masses lies close to the mass of the \KS meson.  This region is excluded as it has high background levels in analyses performed at the threshold experiments.

Finally, the assumption of \CP conservation is removed. Two separate ratios are measured, 
\begin{align}
        R^{+}_{ij} & \equiv \frac{\Gamma[\Dz\to\Kp\pim\pip\pim]}{\Gamma [\Dzb\to\Kp\pim\pip\pim]} \approx r^+_i - r^+_i (\kappa y^{'+})_i \left< \frac{t_{ij}}{\tau} \right> + \frac{({x^+}^2 + {y^+}^2)}{4} \left< \frac{t_{ij}}{\tau}^2 \right>,
\label{eqn:cpv_ratio2}
\end{align}
and the \CP-conjugated observable
\begin{align}
        R^{-}_{ij} & \equiv \frac{\Gamma[\Dzb\to\Km\pip\pim\pip]}{\Gamma [\Dz\to\Km\pip\pim\pip]} \approx r^-_i - r^-_i (\kappa y^{'-})_i \left< \frac{t_{ij}}{\tau} \right> + \frac{( {x^-}^2 + {y^-}^2 )}{4} \left< \frac{t_{ij}}{\tau}^2 \right>,
\label{eqn:cpv_ratio3}
\end{align}
where the different superscripts between the observables in the two expressions allow for the presence of \CP violation.  Only an analysis binned in the phase space is performed, as this provides the best sensitivity.

\section{LHCb detector}
\label{sec:Detector}

The \lhcb detector~\cite{LHCb-DP-2008-001,LHCb-DP-2014-002} is a single-arm forward
spectrometer covering the \mbox{pseudorapidity} range $2<\eta <5$,
designed for the study of particles containing \bquark or \cquark
quarks. The detector used to collect data for this analysis includes a high-precision tracking system
consisting of a silicon-strip vertex detector surrounding the $pp$
interaction region~\cite{LHCb-DP-2014-001}, a large-area silicon-strip detector located
upstream of a dipole magnet with a bending power of about
$4{\mathrm{\,T\,m}}$, and three stations of silicon-strip detectors and straw
drift tubes~\cite{LHCb-DP-2017-001}
placed downstream of the magnet.
The tracking system provides a measurement of the momentum, \ptot, of charged particles with
a relative uncertainty that varies from 0.5\% at low momentum to 1.0\% at 200\gevc.
The minimum distance of a track to a primary $pp$ collision vertex (PV), the impact parameter (IP), 
is measured with a resolution of $(15+29/\pt)\mum$,
where \pt is the component of the momentum transverse to the beam, in\,\gevc.
Different types of charged hadrons are distinguished using information
from two ring-imaging Cherenkov detectors~\cite{LHCb-DP-2012-003}. 
Photons, electrons and hadrons are identified by a calorimeter system consisting of
scintillating-pad and preshower detectors, an electromagnetic
and a hadronic calorimeter. Muons are identified by a
system composed of alternating layers of iron and multiwire
proportional chambers~\cite{LHCb-DP-2012-002}.

The online event selection is performed by a trigger~\cite{LHCb-DP-2012-004}, which consists of a hardware stage, based on information from the calorimeter and muon systems, followed by a two-level software stage, which accesses full-event information at the second level.
In between the two software stages, an alignment and calibration of the detector is performed in near real-time and their results are used in the trigger~\cite{LHCb-PROC-2015-011}.
The same alignment and calibration information is propagated to the offline reconstruction, ensuring consistent and high-quality particle identification (PID) information between the trigger and offline software. 
The identical performance of the online and offline reconstruction offers the opportunity to perform physics analyses directly using candidates reconstructed in the trigger \cite{LHCb-DP-2012-004,LHCb-DP-2016-001}, which the present analysis exploits. 

In the offline selection, trigger signals are associated with reconstructed particles.
Requirements can therefore be made on the trigger selection itself
and on whether the decision was due to the signal candidate, other particles produced in the $pp$ collision, or a combination of both.

Simulation is required to model the effects of the detector acceptance and the imposed selection requirements.
In the simulation, $pp$ collisions are generated using \pythia~\cite{Sjostrand:2007gs,Sjostrand:2006za} with a specific \lhcb configuration~\cite{LHCb-PROC-2010-056}.
Decays of unstable particles are described by \evtgen~\cite{Lange:2001uf}.
For the signal decays, two sets of simulated events are generated using amplitude models fitted to time-integrated RS and WS samples, respectively~\cite{LHCb-PAPER-2017-040}.
The interaction of the generated particles with the detector, and its response, are implemented using the \geant toolkit~\cite{Allison:2006ve, Agostinelli:2002hh} as described in Ref.~\cite{LHCb-PROC-2011-006}. 
The underlying $pp$ interaction is reused multiple times, with an independently generated signal decay for each~\cite{LHCb-DP-2018-004}. The simulated samples are corrected to account for known data-simulation differences in the $\Dstarp$ production kinematics.

The magnetic field deflects oppositely charged particles in opposite
directions and this can lead to detection asymmetries. Periodically
reversing the magnetic field polarity throughout the data-taking almost cancels
the effect. The configuration with the magnetic field pointing upwards (downwards) bends positively (negatively) charged particles
in the horizontal plane towards the centre of the LHC ring.

\section{Signal selection and yield determination}
\label{sec:selection}

At the level of the hardware trigger, events are retained if there is a hadron from the decay of a signal candidate that is above a \pt threshold of 3.5\gevc in the calorimeter, or there is a lepton, photon or hadron with high \pt elsewhere in the event. 
At the first level of the software trigger, events are retained if at least one (or two) tracks from the \Dz decay satisfy the selection criteria of the single-track (two-track) trigger. The single-track trigger requires at least one track with high \pt and high IP significance with respect to all primary vertices in the event.  The two-track trigger requires that a pair of high-\pt tracks forms a  vertex that is significantly displaced from its associated PV, defined as the PV to which the IP significance of the two-track combination is smallest.  A boosted decision tree (BDT) classifier is applied~\cite{Breiman,Roe_2005}, which takes as input the $\chi^2$ of the two-track vertex fit, the number of tracks with high IP significance with respect to their associated PV, the significance of the flight distance of the two-track combination from the associated PV and the summed \pt of the combination~\cite{BBDT,LHCb-PROC-2015-018}.
More information on the typical values of the high-\pt thresholds and other requirements imposed in the trigger may be found in Ref.~\cite{LHCb-DP-2019-001}.
 The full decay chain is reconstructed at the second level of the software trigger, with requirements that are in some cases tightened in the offline selection.  The most important aspects of the full-event reconstruction are now described.

The \Dz candidate is built from four tracks of the appropriate charges and PID hypotheses. The \Dz candidate is then paired with an additional track, referred to as the {\it soft pion}, to construct the \Dstarp candidate. The correlation between the charge of the soft pion and that of the kaon from the \Dz decay allows the candidate to be classified as an RS or WS decay. The four tracks forming the \Dz are required to each have $\pt>250\mevc$, separation from the PV 
and undergo a kinematic constraint to determine their vertex of origin. The soft pion is required to have $\pt > 100 \,(200)\mevc$ for the 2015--16 (2017--18) data set.
Candidates are retained if the \Dz mass falls within the interval $[1840.83,1888.83]\mevcc$, which corresponds to a window of approximately $\pm 3$ times the mass resolution.
A kinematic fit is used to improve the resolution on $\Delta m$, the difference in the reconstructed masses of the \Dstarp and \Dz candidates, and the four-vectors of the five final-state particles.  The fit constrains the mass of the \Dz to its known value~\cite{PDG2024}, and the \Dstarp to decay at the PV. For candidates to be kept for subsequent analysis, it is required that $\Delta m$ lies within $[140,152]\mevcc$.

The dominant background arises from true \Dz candidates paired with a soft pion candidate that does not originate from the signal \Dstarp decay. Additionally, fake \Dz candidates can be formed by combining unassociated hadron tracks. These backgrounds are suppressed by a dedicated BDT classifier.  The BDT is trained on simulated RS decays as a signal proxy, and RS data from the upper $\Delta m$ sideband as a background proxy, using variables associated with the origin and decay vertices and \pt of the \Dz meson, together with the  \chisqip and PID information of the soft pion, where \chisqip\ is defined as 
the difference in the vertex-fit \chisq of the PV reconstructed with and
without the considered track. The BDT threshold is optimised by maximising $S/\sqrt{S+B}$, where $S$ and $B$ are the signal and background yields, respectively, as determined by the fit model described below.

Specific backgrounds are also suppressed using further requirements. 
{\it Secondary} \Dz mesons originating from $b$-hadron decays are suppressed through an upper limit of eight \Dz mean lifetimes, and rejecting candidates with a large \chisqip determined with the \Dz direction of flight, that therefore do not point to the PV. The fraction of these secondary decays remaining in the signal sample after these requirements, $f_{ij}^{\rm sec}$,  is determined from a fit to the \chisqip distribution of \Dz candidates in data for each phase-space bin $i$ and time bin $j$, and is found to increase from around $2\%$ at low decay times to around $8\%$ at five \Dz mean lifetimes, with negligible dependence on the phase-space bin.
The residual contamination from secondaries is accounted for in the mixing measurement by defining the expected WS/RS ratio to be
\begin{align}
    \hat{R}_{ij} = R_{ij} + f^{\rm sec}_{ij} \left( -r_{i}(\kappa y')_{i} \left( \left< \frac{t^{\rm sec}_{ij}}{\tau}\right> - \left< \frac{t_{ij}}{\tau}\right> \right) + \frac{x^2+y^2}{4} \left( \left< \frac{(t^{\rm sec}_{ij})}{\tau}^2 \right> -  \left< \frac{t_{ij}}{\tau}^2\right> \right)  \right),
    \label{eqn:theory_withsec}
\end{align}
where $t^{\rm sec}$ is the true \Dz decay time of secondary events in the sample, as estimated from simulation. The expected ratios  
$\hat{R}^{\pm}_{ij}$ of the $\CP$-violating analysis are defined in an analogous manner.

An RS candidate can be reconstructed as a WS candidate if 
\textit{double misidentification} (DMI) occurs, when the \Km from the \Dz decay is wrongly classified as a pion, and a \pip is wrongly classified as a kaon.
This background, and contamination from \Dz\to\Kp\Km\pip\pim and \Dz\to\pip\pim\pip\pim decays, is suppressed by imposing tight requirements on the PID hypothesis of the kaon and opposite-sign pions from the \Dz decay.  
Candidates in the $m(\Dz)$ sidebands are used to measure the number of these decays remaining in the sample, which is interpolated into the signal $m(\Dz)$ region. The number of DMI background decays is proportional to the number of RS candidates in each phase-space bin, and is subtracted from the measured WS signal yield, where it constitutes a 3\% contamination, with low variation over phase space.

The decays \Dz\to\KS\Kpm\pimp, \KS\to\pip\pim have the same final state as the signal, but occur at different relative rates for WS and RS decays. The contribution of these decays to the signal sample is suppressed through a requirement on the four-track fit quality of the \Dz candidate, which is poor due to the relatively long lifetime of the \KS meson. Additionally, in the inclusive phase-space analysis, candidates are removed if the mass of two opposite-sign pions is within $\pm10\mevcc$ of the known \KS mass~\cite{PDG2024}. In the binned phase-space analysis, these candidates are already excluded through the definition of the bins.
The residual background from this source in the WS sample is estimated from a fit to the mass of the pair of  opposite-sign pions, prior to the removal of candidates in the \KS region, and is found to be 0.07\%. 

Simulation studies indicate that partially reconstructed decays of high-multiplicity final states, such as \Dz\to\Km\pip\pim\pip\piz, do not contaminate the signal region in $m(\Dz)$.  It is possible for the decay mode $\Dz\to\Km\pip\eta', \eta'\to\pip\pim\gamma$ to leak into this region, but at a level that is less than 0.03\% and is therefore neglected.

{\it Clones} occur when two tracks are reconstructed from hits from a single particle.  Candidates constructed from clones possess small angles between pairs of tracks, when considering both the soft pion and the \Dz decay products.  Clone candidates contaminate both the RS and WS samples at a rate of around 2\% and are eliminated from the final samples by demanding a lower bound of $0.03^\circ$ on the opening angle between tracks in the decay. 

A potentially dangerous source of background occurs when the track taken to be the soft pion is reconstructed from hits in the vertex detector that are wrongly matched with hits from the other detectors of the tracking system. The direction of these \textit{ghost tracks} and the $\Delta m$ of the \Dstarp candidates are approximately correct, but the charge assignment can be wrong, 
leading to a migration of true RS decays into the WS sample.
In the track reconstruction, tracks are assigned a probability of being a ghost.  Candidates
containing a soft-pion track with a ghost probability greater than 5\% are removed.
Furthermore, 
in events where more than one WS candidate is reconstructed, only one candidate, chosen arbitrarily, is retained. The same procedure is applied for multiple RS candidates.
If both a WS and RS candidate are reconstructed, the WS candidate is discarded, as this subset of candidates is dominated by RS decays with a wrongly assigned slow pion.
However, analysis of the $\Delta m$ spectrum for the rejected candidates indicates that $(0.78\pm0.06)\%$ of the WS signal is also removed by this procedure.
The measured WS yields are thus corrected for this loss.
Additionally, fiducial requirements, similar to those employed in Ref.~\cite{LHCb-PAPER-2019-006}, are applied based on the momentum vector of the slow pion.  These requirements eliminate regions of high-charge asymmetry and reduce potential biases in the \CP-violation measurement. 

The signal yields are isolated from the remaining background of fake-\Dz and fake-\Dstarp candidates through a binned extended maximum-likelihood fit to the $\Delta m$ distribution. The fit is performed in bins of phase space, time and kaon charge,  simultaneously between WS and RS samples. The RS sample is very pure and so provides information about the signal shape, which is shared between the WS and RS fits. A phase-space inclusive fit is also performed.  
The signal probability density function is empirically described by a modified bifurcated Gaussian with asymmetric tails,
\begin{equation}
\mathcal{P}_\text{sig} (\dm; \mu, \beta, \sigma_L, \sigma_R, \alpha_L, \alpha_R ) = \left\{
\begin{array}{ c l }
\exp \left( \frac{-(\dm -\mu)^2 (1 + \beta (\dm -\mu)^2 )}{2\sigma_L^2 + \alpha_L(\dm-\mu)^2} \right) & \quad \text{for } \dm < \mu \, ,\\
\exp \left( \frac{-(\dm -\mu)^2 (1 + \beta (\dm -\mu)^2 )}{2\sigma_R^2 + \alpha_R(\dm-\mu)^2} \right) & \quad \text{for } \dm \geq \mu \, ,
\end{array}
\right.
\label{eq:cruijff}
\end{equation}
where $\mu$ can be interpreted as the value of \dm at the peak, 
$\sigma_L$ and $\alpha_L$ ($\sigma_R$ and $\alpha_R$) control the width and tail to the left (right) of the peak position, and $\beta$ determines the behaviour of the tails at large off-peak values. All parameters, apart from $\beta$, are free to vary in fits to each time and phase-space bin.
The value of $\beta$ is fixed to a value measured in fits to phase-space integrated signal-only simulated candidates to improve fit stability.
The shape of the background is allowed to be different for each sample, although the same parametrisation is used, which is 
\begin{equation}
\mathcal{P}_\text{bkg} (\dm; \dm_0, a, b) = \left(\dm - \dm_0\right)^{\frac{1}{2}} \left( 1 + a \left(\dm - \dm_0\right) + b \left(\dm - \dm_0\right)^2 \right),
\label{eq:bckgddistrib}
\end{equation}
where $\Delta m_0$ is a low-mass threshold set to the pion mass as  139.57\mevcc, and $a$ and $b$ are free parameters that are fitted in each time and phase-space bin.

\begin{figure}[htb]
\begin{center}
  \includegraphics[width=0.495\textwidth]{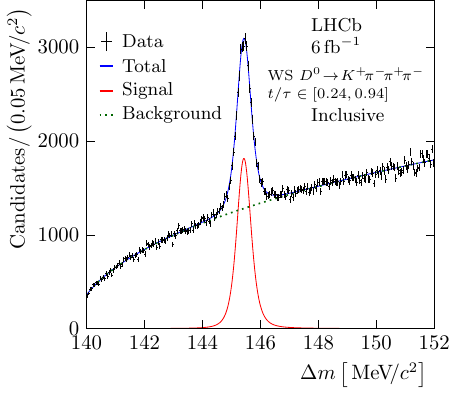}
  \includegraphics[width=0.495\textwidth]{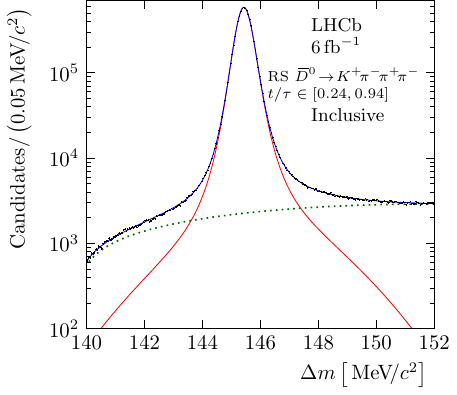}
\caption[Mass fit (\Kp)]{Simultaneous $\Delta m$ fit to phase-space integrated (left) WS  data and (right) RS  data in the first time bin. 
}
\label{fig:data_massfit}
\end{center}
\end{figure}

Example fit results are shown in Fig.~\ref{fig:data_massfit}. The fitted yields, split by WS/RS category and by kaon charge, are presented in Table~\ref{tab:yields}.
Studies performed with simulated pseudoexperiments indicate that the fit procedure is unbiased.

\begin{table}[htb]
\centering
\caption{Signal yields for RS and WS samples, split by kaon charge and integrated over time bins. The results are shown for the inclusive decay and in the individual phase-space bins, where the inclusive and binned results are obtained independently.}
\label{tab:yields}

\begin{tabular}{lrrrr} 
\toprule
 Bin
 & \multicolumn{1}{c}{RS \Km} & \multicolumn{1}{c}{WS \Kp} & \multicolumn{1}{c}{RS \Kp} & \multicolumn{1}{c}{WS \Km} \\ \midrule
Inclusive & $(64\,788\pm9)\times10^3$ & $221\,440\pm830$ & $(62\,802 \pm 9)\times10^3$ & $220\,450\pm820$\\
1 & $(16\,651\pm5)\times10^3$ & $61\,790\pm420$ & $(16\,140\pm5)\times10^3$ & $60\,970\pm420$\\
2 & $(14\,403\pm4)\times10^3$ & $62\,040\pm410$ & $(13\,970\pm4)\times10^3$ & $62\,200\pm410$\\
3 & $(14\,749\pm4)\times10^3$ & $56\,440\pm400$ & $(14\,296\pm4)\times10^3$ & $56\,316\pm400$ \\
4 & $(18\,988\pm5)\times10^3$ & $41\,060\pm410$& $(18\,397\pm5)\times10^3$ & $41\,470\pm410$\\ 
\bottomrule
\end{tabular}
\end{table}

\section{Efficiency corrections}
\label{sec:efficiency}

Different resonant structures populate the WS and RS decays~\cite{LHCb-PAPER-2017-040}. Furthermore, the acceptance of the selection is not uniform over phase space. 
For these reasons, there is a difference in selection efficiency between  WS and RS decays. 
Simulation is used to assess the size of these effects. A weighting is applied to ensure the kinematical distributions  agree between simulation and data; this weighting is derived with the RS samples and applied to both the RS and WS simulation samples.
It is found that the efficiency difference varies from around 0.5\% in phase-space bin~4 to around 2\% in phase-space bin~1.  
Corrections are applied to the measured WS/RS ratios to account for these differences.

Time-dependent efficiency biases also exist and are corrected using simulation. These biases arise from correlations between the acceptance in phase space and decay time that enter at the first stage of the software trigger, where a requirement is imposed on the \pt and $\chisqip$ of tracks from the \Dz decay. In simulated samples generated without mixing, it is found that these correlations lead to a linear evolution in the ratio between WS and RS yields with decay time of approximately 10\% of the size of the genuine mixing signal that is ultimately measured in the phase-space-inclusive data sample.  
With mixing present, the resonant structure of the WS sample varies with decay time, on account of the evolving interference between the DCS and CF amplitudes.  This variation leads to small changes in the correlation that must also be accounted for.   Adjustments are applied in each bin of     decay time to correct for these effects.

In the measurement of the \CP-violation parameters, the quantity of interest is the ratio of partial widths  $R^{\pm}_{ij}$, defined in Eqs.~\ref{eqn:cpv_ratio2} and~\ref{eqn:cpv_ratio3}, for each phase-space bin $i$ and proper-time bin $j$.
Experimentally, however, this ratio can be biased by a charge-detection asymmetry for the soft pion, and a production asymmetry between \Dstarp and $\D^{\ast -}$ mesons.
The expected ratio, 
\robsobsp (\robsobsm),
between the number of WS to RS decays involving a positive (negative) kaon is related to \robsp (\robsm) through
\begin{equation}
\begin{split}
    \robsobs &= 
    \frac{ \int_{ij} \Gamma[\Dz (\Dzb) \to\Kpm\pimp\pip\pim] \epsilon(\Kpm\pimp \pip\pim ) \epsilon(\pis^{\pm})(1 \pm {A_P})}{ \int_{ij} \Gamma[\Dzb (\Dz) \to\Kpm\pimp\pip\pim] \epsilon(\Kpm\pimp \pip\pim ) \epsilon(\pis^{\mp})(1 \mp {A_P})} \\ 
    &\approx \robsprime [ 1 \pm 2( {A_{D}(\pis)}_{ij} + {A_P}_{ij})],
   \end{split}
\label{eq:prodanddetasymm}
\end{equation}
where $\epsilon(\Kpm\pimp \pip\pim )$ is the detection efficiency of the given final state within the phase-space bin, $\epsilon(\pis^{\mp})$ is the detection efficiency of the slow pion of the indicated charge, ${A_P}_{ij}$ is the production asymmetry, ${A_D(\pis)}_{ij}$ is the detection asymmetry of the soft pion, \robsprime are the ratios corrected for contamination from secondary decays 
as in Eq.~\ref{eqn:theory_withsec},
and the integration is over the phase-space and decay-time bin.
In principle, neither the detection nor the production efficiency has a dependence on the time of the decay, but such an effect can enter through correlations in the selection. 
The approximate equality in Eq.~\ref{eq:prodanddetasymm} is valid when the detection and production asymmetries are small, which is the case. 

The values of the sum ${A_{D}(\pis)}_{ij} + {A_P}_{ij}$ are determined from data using a control sample of $\Dstarp\to\Dz\pip$, $\Dz\to\Kp\Km$ decays.  Here, the observed $\CP$ asymmetry in each phase-space and decay-time bin is 
\begin{equation}
   A^{KK}_{ij} =   {A_{D} (\pis)}_{ij} + {A_{P}}_{ij} + ( a^{d}_{KK} + \left< \frac{t_{ij}}{\tau} \right> \Delta Y ) \, , 
   \label{eq:cpvkk}
\end{equation}
where $a^{d}_{KK}$ is the \CP asymmetry in decay and $\Delta Y$ is the time-dependent \CP asymmetry of the decay, both of which have been measured by LHCb and determined to be \mbox{($7.7\pm5.7)\times10^{-4}$}~\cite{LHCb-PAPER-2022-024} and $(1.0\pm1.1)\times10^{-4}$~\cite{LHCb-PAPER-2020-045}, respectively.  

The control sample is selected using the same requirements on the trigger and the key variables as the signal decay. 
A weighting is applied in each phase-space and decay-time bin, based on the kinematics of the \Dz and soft-pion candidates, to ensure that the results obtained from the control channel are applicable to the signal sample.

The asymmetries $A^{KK}_{ij}$ are determined from a fit to the \dm distribution, which is parameterised in a similar way to that of the signal sample. 
The \Dz and \Dzb samples are fitted simultaneously, with the asymmetries extracted directly. 
Fits are performed for each data-taking period, magnet polarity, phase-space bin and decay-time bin.
The asymmetries are averaged over each data-taking period, weighted by the number of signal candidates from the period. 

The typical sizes of ${A_{D}(\pis)}_{ij} + {A_P}_{ij}$  are found to be in the range of 0.5--1.0\% with a precision of $\sim$0.15\%. 
When including these corrections in the fit to measure the \CP-violation parameters, the values of $A^{KK}_{ij}$, $a^{d}_{KK}$, and $\Delta Y$ are constrained within their uncertainties using Gaussian priors. 

\FloatBarrier

\section{Fit to mixing parameters}
\label{sec:mixingfit}

\begin{figure}[b]
\begin{center}
\includegraphics[width=0.495\textwidth]{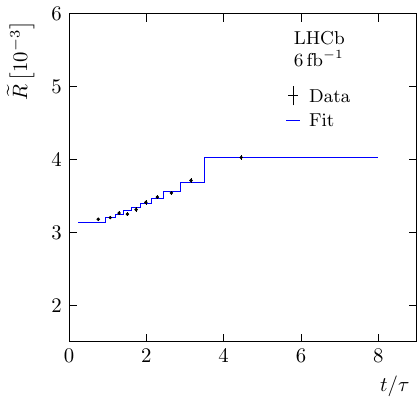}
\vspace*{-0.5cm}
\end{center}
\caption[Inclusive fit to data]{Inclusive WS/RS ratio in bins of decay time, with the fit result also shown. }
\label{fig:cpint_mix_results_inclusive}
\end{figure}

The WS/RS ratio in data, $\widetilde{R}_{ij}$, is constructed by dividing the measured WS signal yield by the RS signal yield, after subtracting DMI background in both cases. 
This quantity can be determined separately for the WS samples with positive and negative kaons. The fit takes as input the geometric mean of these two ratios.

The mixing fit minimises the metric 
\begin{align}
\chisq = \sum_{ij} \left( \frac{\widetilde{R}_{ij} - \hat{R}_{ij}}{\sigma_{\widetilde{R}_{ij}}}\right)^2 \, ,
\end{align}
where $\hat{R}_{ij}$ is the expected WS/RS ratio, in the presence of contamination from secondary decays, defined in Eq.~\ref{eqn:theory_withsec}, and $\sigma_{\widetilde{R}_{ij}}$ is the uncertainty on the measured ratio. In the inclusive fit, the three parameters $r$, $(\kappa y')$ and $x^2+y^2$ vary freely.   
In the phase-space binned fit there are nine free parameters:  $r_i$, $(\kappa y')_i$ and $x^2+y^2$. 

\begin{figure}[htb]
\begin{center}
\includegraphics[width=0.85\textwidth]{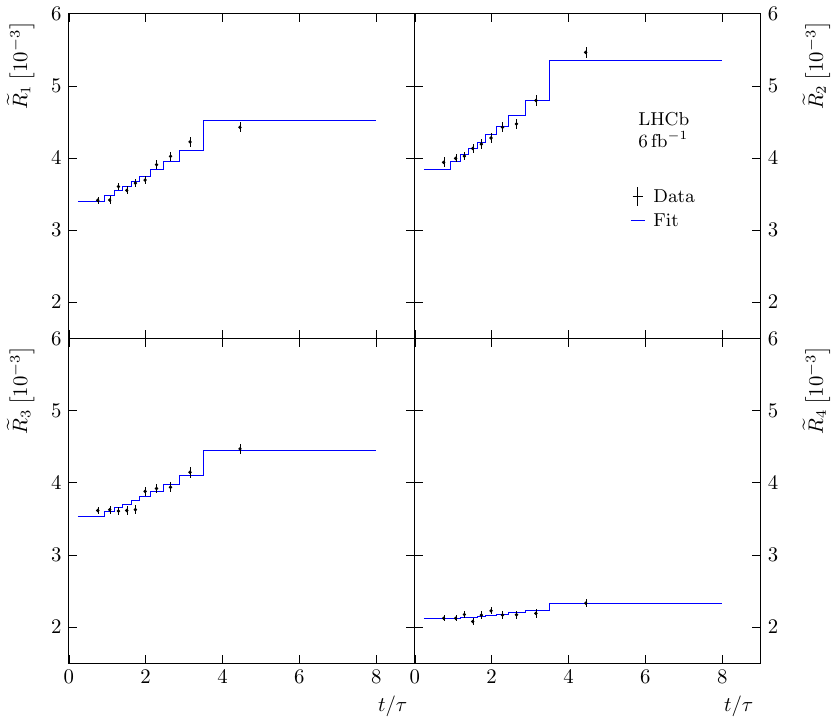}
\vspace*{-0.5cm}
\end{center}
\caption[Inclusive and binned fit to data]{Phase-space binned WS/RS ratio in bins of decay time, with the fit results also shown. }
\label{fig:cpint_mix_results_binned}
\end{figure}

\begin{figure}[htb]
\begin{center}
\includegraphics[width=0.85\textwidth]{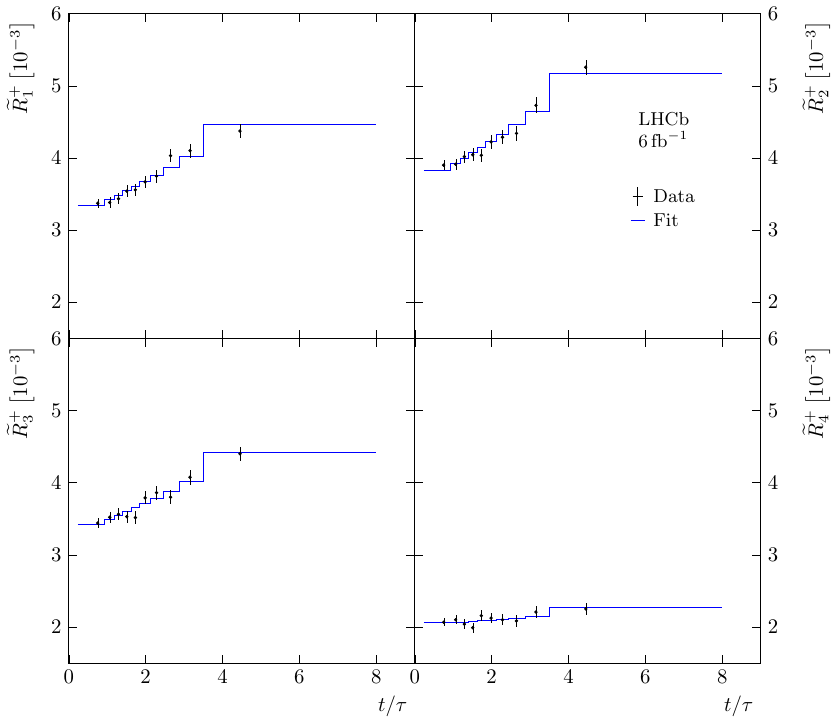}
  
\includegraphics[width=0.85\textwidth]{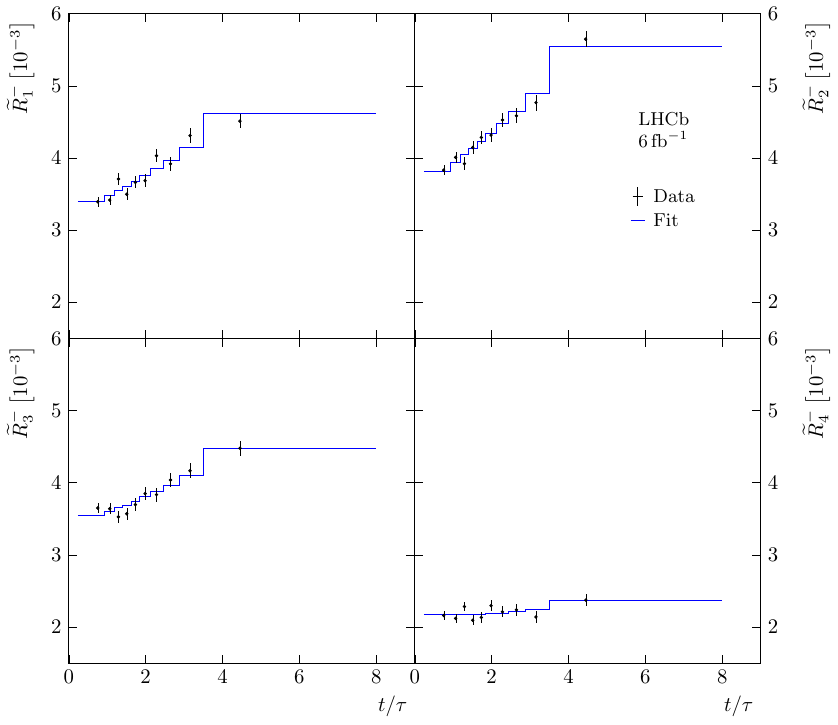}

\vspace*{-0.5cm}
\end{center}
\caption[Phase-space binned fit to data]{Phase-space binned (top) $\Tilde{R}^{+}$ and (bottom)~$\Tilde{R}^{-}$ WS/RS ratio in bins of decay time, with  the fit results also shown.}
\label{fig:cpv_mix_results}
\end{figure}

The fits to data are shown in Figs.~\ref{fig:cpint_mix_results_inclusive} and~\ref{fig:cpint_mix_results_binned} and the numerical results in Table~\ref{tab:mixingresults}. The reduced $\chisq$ for the inclusive fit is 1.07 for 7 degrees of freedom, with the corresponding numbers being 1.18 and 31 for the phase-space binned fit. The inclusive fit manifests a distinctive near-linear slope, which is characteristic of the mixing effects predicted by Eq.~\ref{eqn:mixing} when the values of $x$ and $y$ are small.  The behaviour, though qualitatively similar, varies between the phase-space bins, which is expected if the mean phase difference between \Dz and \Dzb decays, or the coherence factor, is not the same in each bin. 
Studies performed with large ensembles of pseudoexperiments demonstrate that the fit procedure is unbiased.

To measure the \CP-violation parameters, the two $\widetilde{R}^{\pm}_{ij}$ ratios are independent inputs to the fit, along with the asymmetry corrections described in Sec.~\ref{sec:efficiency}. 
In this case, only a phase-space binned fit is performed. 
Eighteen mixing parameters, $r_i^\pm$, $(\kappa y'^\pm)_i$ and 
$\left( x^{\pm\,2} + y^{\pm\,2} \right)$,
vary freely in the fit.  In addition, there are 42 parameters associated with the asymmetry correction that are constrained within uncertainties to their measured values. These are the set of 40 per-bin asymmetries $A^{KK}_{ij}$ and the two parameters characterising \CP violation in the control channel with which they are measured: $a^{d}_{KK}$ and $\Delta Y$.
The fits to data are shown in Fig.~\ref{fig:cpv_mix_results} and the numerical results in Table~\ref{tab:mixingresults_cpv}. The reduced $\chisq$  is 0.86 for 71 degrees of freedom. 

\begin{table}
  \caption[Final results]{
        Measured values of the mixing parameters in the \CP-conserving fit. 
        }
        \vspace*{-0.5cm}
\begin{center}
\renewcommand{\arraystretch}{1.2}
\begin{tabular}{l       c    r @{\:$\pm$\:} l @{\:$\pm$\:} l         r @{\:$\pm$\:} l @{\:$\pm$\:} l  }
    \toprule
    Phase-space bin      & \multicolumn{1}{c}{$(x^2 + y^2)$ $[\times 10^{-5}]$} & \multicolumn{3}{c}{$r_{(i)}$ $[\times 10^{-2}]$}  & \multicolumn{3}{c}{$(\kappa y')_{(i)}$ $[\times 10^{-3}]$}  \\ 
    \midrule 
    Inclusive                 & $7.2 \pm 2.8 \pm 1.6$     & $5.511 $ & $0.035 $ & $0.026$ & $-2.69 $ & $0.67 $ & $0.39$ \\
    1  & \multirow{4}{*}{$5.9 \pm 2.8 \pm 1.2$}      & $5.674 $ & $0.045 $ & $0.029$ & $-4.16 $ & $0.73 $ & $0.30$ \\ 
    2  &                                             & $5.989 $ & $0.049 $ & $0.031$ & $-5.72 $ & $0.75 $ & $0.36$ \\ 
    3  &                                             & $5.834 $ & $0.046 $ & $0.032$ & $-3.01 $ & $0.72 $ & $0.36$ \\ 
    4  &                                             & $4.627 $ & $0.051 $ & $0.043$ & $0.51 $ & $0.83 $ & $0.40$ \\ 
    \bottomrule
  \end{tabular}
  \end{center}
\label{tab:mixingresults}
\end{table}

\begin{table}
  \caption[Final results]{
        Measured values of the mixing parameters in the \CP-violating fit.}
        \vspace*{-0.5cm}
\begin{center}
\renewcommand{\arraystretch}{1.2}
\begin{tabular}{l       c    r @{\:$\pm$\:} l @{\:$\pm$\:} l         r @{\:$\pm$\:} l @{\:$\pm$\:} l  }
    \toprule
    Phase-space bin      & \multicolumn{1}{c}{$({x^+}^2 + {y^+}^2) $ $[\times 10^{-5}]$} & \multicolumn{3}{c}{$r^+_{i}$ $[\times 10^{-2}]$}  & \multicolumn{3}{c}{$(\kappa y^{'+})_i$ $[\times 10^{-3}]$}  \\ 
    \midrule 
1  & \multirow{4}{*}{$6.1$ $\pm$ $4.2$ $\pm$ $1.3$}  
& $5.662$ &  $0.066$ & $0.029$  & $-4.16$ &  $1.08$ & $0.29$  \\ 
2  &                                     
& $6.036$ &  $0.066$ & $0.030$  & $-4.97$ &  $1.05$ & $0.32$ \\ 
3  &                                     
& $5.762$ &  $0.068$ & $0.031$  & $-3.39$ &  $1.08$ & $0.34$  \\ 
4  &                                     
& $4.590$ &  $0.075$ & $0.044$  & $0.51$ &  $1.24$ & $0.41$ \\ \bottomrule \toprule
    Phase-space bin      & \multicolumn{1}{c}{$({x^-}^2 + {y^-}^2)$ $[\times 10^{-5}]$} & \multicolumn{3}{c}{$r^-_{i}$ $[\times 10^{-2}]$}  & \multicolumn{3}{c}{$(\kappa y^{'-})_i$ $[\times 10^{-3}]$}  \\ 
    \midrule 
1  & \multirow{4}{*}{$6.8$ $\pm$ $4.2$ $\pm$ $1.3$} 
& $5.625$ &  $0.066$ & $0.029$  & $-4.30$ &  $1.08$ & $0.29$\\ 
2  &                                     
& $5.884$ &  $0.067$ & $0.030$  & $-6.53$ &  $1.08$ & $0.32$  \\ 
3  &                                     
& $5.794$ &  $0.067$ & $0.031$  & $-2.87$ &  $1.06$ & $0.34$   \\ 
4  &                                     
& $4.648$ &  $0.074$ & $0.044$  & $0.84$ &  $1.21$ & $0.42$  \\ 
    \bottomrule
  \end{tabular}
  \end{center}
\label{tab:mixingresults_cpv}
\end{table}

\FloatBarrier

\section{Systematic uncertainties}
\label{sec:systematics}

Systematic uncertainties are assigned on the parameters determined from the mixing fits. These are associated with the detector resolution, the understanding of residual backgrounds, the determination of the signal yields, and the efficiency corrections. The systematic uncertainties in the \CP-conserving fits are summarised in Tables~\ref{tab:syst_unc_cpint_inclusive} and~\ref{tab:syst_unc_cpint_binned}.  The uncertainties for the \CP-violating fit are shown in Table~\ref{tab:all_syst_psbinned_CPV}. 

There is a bias arising from the finite resolution on the decay time and the location of the decay in phase space.  This is studied through fits to many simulated data samples, generated with known mixing parameters. The observed bias on each fit parameter is assigned as a systematic uncertainty. 

The uncertainty associated with the treatment of secondary decays in the sample is assessed by setting the expected fraction of these decays to zero in the mixing fit, and assigning the change in fitted values as the corresponding systematic uncertainty. An alternative approach, in which the expected secondary fraction is doubled in the fit, returns changes that are similar in magnitude.

The uncertainty on the level of DMI contamination is determined by adding in quadrature the uncertainty on the background from the baseline analysis to the variations observed when different fit models are used to measure this contribution.  Many samples are then generated from the data set, each with the level of subtracted DMI background taken from a Gaussian function of width set to the assigned uncertainty. The spread in fit results from this exercise is taken as the corresponding systematic uncertainty for each parameter.

The correction factor applied to account for the loss in WS signal for events where the WS candidate has an overlapping RS decay is known with limited precision. To assess the associated uncertainty, the fit is repeated many times with the correction factor fluctuated within its uncertainties. The spread of the resulting fitted parameters is used to assign an uncertainty.  

The amount of residual background from slow-pion ghosts is assessed by examining the stability of the WS/RS ratio with the ghost-probability variable. The bias on the inclusive samples is found to be $(0.1\pm0.5)\%$, which is assumed to be uniform across phase space.
Many  pseudosamples are then generated and fit in which the WS contribution in each phase-space bin is adjusted by a common random number sampled from a Gaussian function with parameters assigned from the ghost-probability study.  The spread in results is used to assign the uncertainties on the fit parameters.
This contamination has no dependence on decay time and therefore affects the $r$ parameter more than the other fit variables.
A similar procedure is followed to assign uncertainties associated with  backgrounds from charm decays such as \Dz\to\KS\Kp\pim and $\Dz\to\Kp\pim\eta$, which are found to be small.

The determination of the signal yields is affected by the choice of empirical shapes and fixed parameters used to fit the \dm distributions. 
The signal parametrisation is well determined, as this is fitted from the very large and pure RS samples.
The parameter $\beta$ in Eq.~\ref{eq:cruijff} is fixed from fits to phase-space integrated, signal-only simulated \dm distributions and is varied within its uncertainty to assess the uncertainty due to this choice. 
The largest uncertainty in the determination of the WS yields arises from the choice of 
background parametrisation. The default parametrisation of Eq.~\ref{eq:bckgddistrib} is changed to an alternative description
\begin{equation}
\mathcal{P}_\text{alt bkg} (\dm; \dm_0, a, b) = \exp\left( -b \left(\dm - \dm_0\right)\right) \cdot \left(\dm - \dm_0\right)^{a}, 
\label{eq:altbckgddescrip}
\end{equation}
where $\dm_0$ is the same low-mass threshold as in Eq.~\ref{eq:bckgddistrib}, and $a$ and $b$ are free parameters. 
Pseudoexperiments are performed, where the background distribution is generated according to Eq.~\ref{eq:altbckgddescrip}, with parameters determined from data. The yields of the pseudoexperiments are then fitted with the baseline model, and the shifts in the means of the mixing parameters are taken as the systematic uncertainties.

The largest systematic uncertainty arises from the efficiency corrections.
This uncertainty receives contributions from the size of the simulation samples used to determine the correction, limited  knowledge of the changing interference between the DCS and CF amplitudes, and any imperfections in  agreement  between simulation and data.  It is the size of the simulation samples that dominates, and it is this source that is used to assign the uncertainty.
No additional uncertainties are assigned for the corrections associated with the slow-pion detection and \Dstarpm production asymmetries, as these corrections are constrained fit parameters in the \CP-violation analysis. The contribution to the total uncertainty is around half that coming from the size of the WS sample.  

Finally, for the inclusive analysis there is a bias on the fit parameters arising from the exclusion of a small region of phase space through the removal of \KS contamination.  This bias is assessed by measuring the size of the effect in simulation, which is then assigned as the systematic uncertainty.

\begin{table}[tb]
\centering
\caption[]{
        Systematic uncertainties for the  \CP-conserving inclusive fit. 
        All values are scaled by $10^{5}$. 
        }
\label{tab:syst_unc_cpint_inclusive}
\renewcommand{\arraystretch}{1.2}
\begin{tabular} { l r r r } 
\toprule 
Uncertainty      & $(x^2 + y^2)$  & $r$ & $(\kappa y')$  \\ 
\midrule 
Detector resolution                    & 0.2 & 2.3 & 5.6 \\ 
Secondaries correction              & 0.2 & 2.0 & 8.9 \\ 
DMI correction            & 0.8 & 10.8 & 19.7  \\ 
WS/RS overlap removal            & $<0.1$ & 2.6 & 0.1  \\ 
Slow-pion ghosts                    & $<0.1$ & 13.8 & 0.6 \\ 
Charm backgrounds                   & $<0.1$ & 2.8 & 0.1 \\ 
Signal shape        & $<0.1$ & 0.3 & 0.8 \\
Background shape                      & 0.1 & 5.5 & 0.3 \\ 
Efficiency correction               & 1.3 & 17.5 & 32.2 \\ 
Removal of \KS region              & 0.2 & 5.3 & 2.2 \\ 
\midrule
Total systematic                    & 1.6 & 26 & 39 \\
Statistical                         & 2.8 & 35 & 67 \\
\midrule
Total                               & 3.2 & 44 & 78 \\ 
\bottomrule
\end{tabular} 
\end{table}

\begin{table}
\centering
\caption[]{
        Systematic uncertainties for \CP-conserving phase-space binned fits.
        All values are scaled by $10^{5}$. 
        }
        \label{tab:syst_unc_cpint_binned}
\setlength{\tabcolsep}{4.1pt}
\begin{tabular} { l r r r r r r r r r} \toprule 
\small
Uncertainty     & $(x^2 + y^2)$
& $r_1$ & $r_2$ & $r_3$ & $r_4$ & $(\kappa y')_1$ & $(\kappa y')_2$ & $(\kappa y')_3$ & $(\kappa y')_4$  \\ 
\midrule 
Detector resolution                     & 0.2 & 17.9 & 12.0 & 17.1 & 29.0 & 2.6 & 10.6 & 16.5 & 7.7 \\
Secondaries correction               & 0.2 & 1.6 & 5.9 & 3.1 & 2.3 & 10.0 & 17.3 & 10.6 & 2.0 \\
DMI correction             & 0.8	& 12.8	& 13.7	& 12.5	& 17.3	& 19.7	& 19.9	& 18.7	& 24.8
 \\
 WS/RS overlap removal            & $<0.1$ & 2.7 & 2.8 & 2.8 & 2.2 & 0.2 & 0.3 & 0.1 & $<0.1$ \\
Slow-pion ghosts                     & $<0.1$ & 14.0 & 14.7 & 14.2 & 11.5 & 0.8 & 1.3 & 0.8 & 0.2 \\
Charm backgrounds                   & $<0.1$ & 2.9 & 3.1 & 3.0 & 2.4 & 0.1 & 0.3 & 0.1 & $<0.1$ \\ 
Signal shape        & $<0.1$ & 0.5 & 0.6 & 0.6 & 0.5 & 0.4 & 0.4 & 0.4 & 0.4 \\
Background shape                       & 0.1 & 4.2 & 10.1 & 9.8 & 2.6 & 1.3 & 9.7 & 6.1 & 6.9\\
Efficiency correction                & 0.8 & 12.4 & 16.9 & 16.5 & 24.4 & 19.8 & 19.8 & 22.6 & 29.6 \\
\midrule
Total systematic                     &1.2 & 29 & 31 & 32 & 43 & 30 & 36 & 36 & 40 \\

Statistical                          & 2.8 & 45 & 49 & 46 & 51 & 73 & 75 & 72 & 83 \\
\midrule
Total                               &  3.0  &  54  &  58  &  56  &  67  &  79  &  83  &  80  &  92  \\ 
\bottomrule
\end{tabular} \\
\addtolength{\tabcolsep}{0.5pt}
\end{table}

\begin{table}[htb]
\centering
\caption[All systematic uncertainties for CPV fit]{
        Systematic uncertainties for phase-space binned \CP-violating fits. All values are scaled by $10^{5}$.  
        Note that the asymmetry correction uncertainty is not included in the total systematic as it is incorporated through a constraint into the statistical uncertainty of the fit. All contributions are fully correlated between the $\Tilde{R}^+$ and $\Tilde{R}^-$ entries, apart from the secondary-correction and asymmetry-correction components, which are uncorrelated.
        }
\label{tab:all_syst_psbinned_CPV}
\def\arraystretch{1.2}
\addtolength{\tabcolsep}{-3.8pt}
\begin{tabular} { l r r r r r r r r r} \toprule
$\Tilde{R}^{+}$ uncertainty        & $({x^+}^2\!\!+\! {{y^+}}^2)$  & $r^+_1$ & $r^+_2$ & $r^+_3$ & $r^+_4$ & $(\kappa y^{'+})_1$ & $(\kappa y^{'+})_2$ & $(\kappa y^{'+})_3$ & $(\kappa y^{'+})_4$  \\ 
\midrule 
Detector resolution                     & 0.2 & 17.9 & 12.0 & 17.1 & 29.0 & 2.6 & 10.6 & 16.5 & 7.7 \\ 
Secondaries correction               & 0.4 & 0.6 & 0.5 & 0.5 & 4.0 & 4.0 & 5.1 & 1.6 & 9.0 \\     
DMI correction             & 0.8	& 13.2	& 14.5	& 12.9	& 17.2	& 20.8	& 21.2& 19.8	& 25.2 \\ 
 WS/RS overlap removal       & $<0.1$ & 2.7 & 2.9 & 2.7 & 2.2 & 0.2 & 0.2 & 0.2 & $<0.1$ \\
Slow-pion ghosts                     & $<0.1$ & 14.0 & 14.7 & 14.2 & 11.5 & 0.8 & 1.3 & 0.8 & 0.2 \\
Charm backgrounds                    & $<0.1$ & 2.9 & 3.1 & 3.0 & 2.4 & 0.1 & 0.3 & 0.1 & $<0.1$ \\ 
Signal shape         & $<0.1$ & 0.5 & 0.6 & 0.6 & 0.5 & 0.4 & 0.4 & 0.4 & 0.4 \\
Background shape                       & 0.3 & 0.7 & 1.1 & 0.4 & 1.1 & 0.7 & 4.3 & 3.1 & 4.8\\
Efficiency correction                & 0.8 & 12.4 & 16.9 & 16.5 & 24.4 & 19.8 & 19.8 & 22.6 & 29.6 \\
Asymmetry correction                & 1.9 & 24.2 & 25.9 & 24.9 & 27.9 & 44.4 & 44.1 & 44.5 & 51.7 \\
\midrule
Total systematic   &  1.3 &  29 &  30 &  31 &  44 &  29 &  32 &  34 &  41  \\     
Statistical                          & 4.2 &  66 &  66 &  68 &  75 &  108 &  105 &  108 &  124 \\ 
\midrule
Total                               &   4.4 &  72 &  72 &  75 &  87 &  112 &  109 &  113 &  131 \\ \bottomrule
\toprule
$\Tilde{R}^{-}$ uncertainty        & $({x^-}^2\!\! +\!{{y^-}}^2)$  & $r^-_1$ & $r^-_2$ & $r^-_3$ & $r^-_4$ & $(\kappa y^{'-})_1$ & $(\kappa y^{'-})_2$ & $(\kappa y^{'-})_3$ & $(\kappa y^{'-})_4$  \\ 
\midrule 
Detector resolution                     & 0.2 & 17.9 & 12.0 & 17.1 & 29.0 & 2.6 & 10.6 & 16.5 & 7.7 \\ 
Secondaries correction               & 0.6 & 0.2 & 0.9 & 1.4 & 4.8 & 2.1 & 6.3 & 1.7 & 12.3 \\  
DMI correction             &0.8	& 13.2	& 14.5	& 12.9	& 17.2	& 20.8	& 21.2	& 19.8	& 25.2 \\ 
 WS/RS overlap removal       &  $<0.1$ & 2.7 & 2.8 & 2.8 & 2.2 & 0.2 & 0.3 & 0.1 & $<0.1$ \\
Slow-pion ghosts                     & $<0.1$ & 14.0 & 14.7 & 14.2 & 11.5 & 0.8 & 1.3 & 0.8 & 0.2 \\
Charm backgrounds                    & $<0.1$ & 2.9 & 3.1 & 3.0 & 2.4 & 0.1 & 0.3 & 0.1 & $<0.1$ \\ 
Signal shape         & $<0.1$ & 0.5 & 0.6 & 0.6 & 0.5 & 0.4 & 0.4 & 0.4 & 0.4 \\
Background shape                       & 0.1 & 1.0 & 3.2 & 0.8 & 0.4 & 0.3 & 5.6 & 1.6 & 0.7\\
Efficiency correction                & 0.8 & 12.4 & 16.9 & 16.5 & 24.4 & 19.8 & 19.8 & 22.6 & 29.6 \\
Asymmetry correction                & 1.9 & 24.1 & 24.8 & 24.3 & 27.3 & 44.7 & 44.1 & 43.2 & 51.0 \\
\midrule
Total systematic  & 1.3 &  29 &  30 &  31 &  44 &  29 &  32 &  34 &  42  \\  
Statistical                          &  4.2 &  66 &  67 &  67 &  74 &  108 &  108 &  106 &  121  \\
\midrule
Total                               & 4.4 &  72 &  73 &  74 &  86 &  112 &  113 &  111 &  128 \\ \bottomrule 
\end{tabular} \\
\addtolength{\tabcolsep}{1pt}
\end{table}

In order to validate the robustness of the results, the data are split into subsamples according to the data-taking period, magnet polarity, the hardware and first-level software trigger decision used to select each event, the slow-pion transverse momentum, and the estimated probability that the slow pion is formed from a ghost track. All subsamples yield consistent results.  The largest difference observed is of around two statistical standard deviations, when the data are divided by magnet polarity.

\section{Interpretation}
\label{sec:interpretation}

The measurements of the observables presented in Sec.~\ref{sec:mixingfit} are interpreted with two approaches.
In the first, the allowed region for the hadronic parameters of the $\Dz$-meson decay is determined by constraining the mixing parameters $(x,y)$ to the values given in Ref.~\cite{LHCb-CONF-2024-004}. 
In the second interpretation the mixing parameters $(x,y)$ and the corresponding $\CP$-violating parameters $(\Delta x, \Delta y)$ are determined by including constraints on the hadronic parameters from charm-threshold data.

The parameters for both interpretations are determined by defining the metric  
\begin{equation}
  \chi^2_\text{mix} =  \sum_{ij} [\mathbf{V}^{-1}]_{ij} \left( O_i - \hat{O}_i \right) \left( O_j - \hat{O}_j \right),
\end{equation}
where $O_i$ is the $i$th observable reported in Tables~\ref{tab:mixingresults} and \ref{tab:mixingresults_cpv}. The covariance matrix $\mathbf{V}$ is constructed from the statistical and systematic correlation matrices given in Appendix~\ref{app:stat_syst_cov} and the uncertainties reported in Tables~\ref{tab:syst_unc_cpint_inclusive}, \ref{tab:syst_unc_cpint_binned} and~\ref{tab:all_syst_psbinned_CPV}. 
The expectation values, $\hat{O}$, are expressed in terms of the underlying physics parameters $r_{(i)},\kappa_{(i)},\delta_{(i)},x$ and $y$ for the phase-space inclusive (binned) case, with $\Delta x$ and $\Delta y$ also included when the fit allows for \CP violation.  

The fit minimises the sum of $\chi^2_\text{mix}$ and one or more constraint terms, depending on the purpose of the analysis. The minimisation is made with respect to both the mixing parameters and the hadronic parameters of the $\Dz$ decay. 
In its most complete form, the total $\chi^2$ is given by
\begin{equation}
\chi^2_\text{total} = \chi^2_\text{mix} + \chi^2_\text{charm} + \chi^2_\text{CLEO} +\chi^2_\text{BESIII} + \chi_{
\text{R1}}^2,
\end{equation}
where $\chi^2_\text{charm}$ is the constraint on the charm-mixing parameters, and 
$\chi^2_\text{CLEO} $ and $\chi^2_\text{BESIII}$ are the constraints from charm-threshold data from the CLEO-c~\cite{Evans2020_new} and BESIII experiments~\cite{BESIII2021_new}, respectively. 
The BESIII term is calculated using the supplementary material provided in Ref.~\cite{BESIII2021_new}. 
The term $\chi_{\text{R1}}^2$ constrains the  parameters $\kappa$, $\delta$ and $r$ using the inclusive results from LHCb data collected in Run~1 of the LHC~\cite{LHCb-PAPER-2015-057}. This constraint can be applied in the binned case by expressing the inclusive quantities in terms of the binned ones using the relations
\begin{equation}
  \begin{split}
\kappa e^{i\delta} &= \frac{1}{r} \sum_i K_i r_i \kappa_i e^{i\delta_i}, \\
  r  &= \sqrt{\sum_i K_i r_i^2},
  \end{split}
\end{equation}
where $K_i$ is the fractional yield of CF decays in the $i$th bin~\cite{LHCb-PAPER-2017-040,Evans2020_new}. 

\subsection{Interpretation in terms of hadronic parameters}

\begin{figure}[htb]
\centering 
\includegraphics[width=0.495\textwidth]{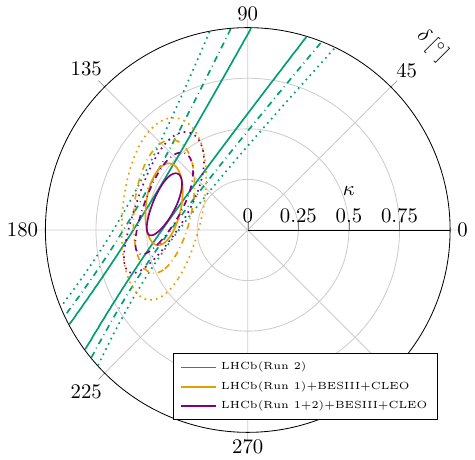}
\caption{\label{fig:rd_global}Profile likelihood of $\kappa$ \vs
$\delta$, inclusive over the entire phase space. The \mbox{$\Delta \chi^2=2.38$}, $~6.18,~11.83$ contours are indicated, corresponding to the $68.7,~95.0,~99.7\%$ confidence intervals for a Gaussian likelihood.}
\end{figure}

The inclusive \CP-conserving mixing observables, given in Table~\ref{tab:mixingresults}, are interpreted in terms of the hadronic parameters of the $D$ decay using external constraints on the charm-mixing parameters from Ref.~\cite{LHCb-CONF-2024-004} through the inclusion of the $\chi^2_\text{charm}$ term in the fit.
The mixing observables are insufficient to give unique values of the hadronic parameters, instead forming bands in the $\kappa, \delta$ likelihood plane, as shown in Fig.~\ref{fig:rd_global}. 
The prior knowledge of these parameters is also indicated from the CLEO-c, BESIII and LHCb Run~1 data~\cite{Evans2020_new,BESIII2021_new,LHCb-PAPER-2015-057}, together with the result of combining previous measurements with the results of the present analysis.  
The compatibility of the different sets of measurements is good, with $\chi^2_\text{mix} = 1.3$ for two degrees of freedom,  
and the inclusion of the results from the current analysis improves the knowledge of the hadronic parameters~\cite{BESIII2021_new}.

\begin{table}
\centering
\caption{\label{table:hadronic_parameters_fit}
        Central values of the hadronic parameters after 
        combination of this measurement with previous results from BESIII~\cite{BESIII2021_new} and CLEO-c~\cite{Evans2020_new} charm-threshold data,  
        and the LHCb Run~1 phase-space-inclusive mixing analysis~\cite{LHCb-PAPER-2015-057}. 
        }
\label{tab:interpret}
    \def\arraystretch{1.2}

\begin{tabular} { l c c c }
\toprule
Phase-space bin   & $r_{(i)} \,[\times 10^{-2}]$ & $\kappa_{(i)}$  & $\delta_{(i)} [^\circ]$  \\
\midrule
Inclusive  & \rgmeas & \kgmeas & \dgmeas \\
Bin 1 & $5.68 \pm 0.04$& $0.598\aerr{0.106}{ 0.088 }$& $123.8\aerr{13.3}{ 10.7 }$\\
Bin 2 & $5.98 \pm 0.04$& $0.777\aerr{0.073}{ 0.064 }$& $149.7\aerr{11.1}{ 10.8 }$\\
Bin 3 & $5.82 \pm 0.04$& $0.663\aerr{0.091}{ 0.087 }$& $196.3\aerr{8.7}{ 11.2 }$\\
Bin 4 & $4.64 \pm 0.05$& $0.075\aerr{0.108}{ 0.049 }$& $319.3\aerr{82.5}{ 58.3 }$\\
\bottomrule
\end{tabular}

\vskip 2em
\end{table}

\begin{figure}[htb]
  \includegraphics[width=0.495\textwidth]{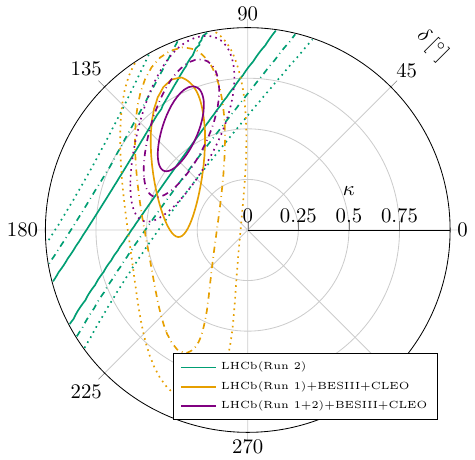}
  \includegraphics[width=0.495\textwidth]{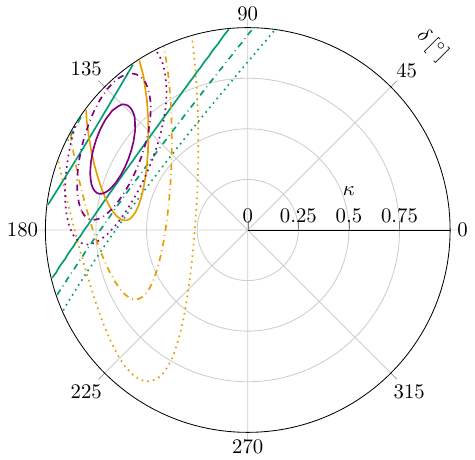}

  \includegraphics[width=0.495\textwidth]{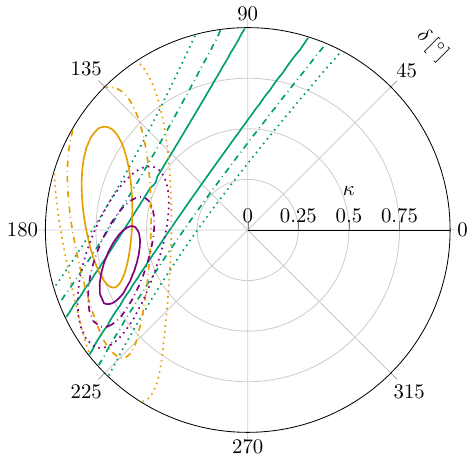}
  \includegraphics[width=0.495\textwidth]{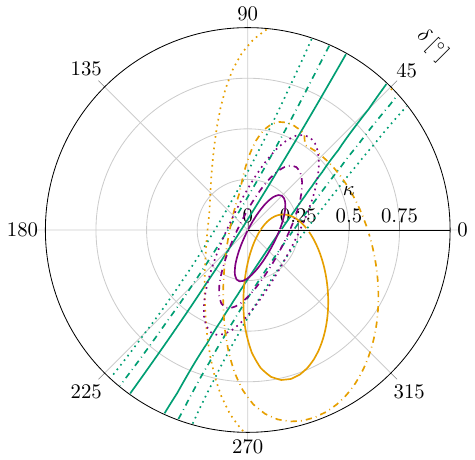}

\caption{\label{fig:rd_local}Likelihood contours of $\kappa$ \vs $\delta$, for each of the bins of $\Dz$ phase space. The \mbox{$\Delta\chi^2=2.38$},$~6.18,~11.83$ contours are indicated, corresponding to the $68.7,~95.0,~99.7\%$ confidence intervals for a Gaussian likelihood.}
\end{figure}

The likelihood contours for the binned hadronic parameters are shown in Fig.~\ref{fig:rd_local}, comparing the bands from the mixing analysis with the allowed regions from charm-threshold data and the phase-space integrated LHCb analysis of Run~1 data, and the combination of the two sets of results.  It is evident that the inclusion of the new charm-mixing results brings significant improvement to the knowledge of the parameters in each bin.
The compatibility of the measurements is good, with the $\chi_\text{mix}^2$ being 0.3 for 5 degrees of freedom, which assumes that the hadronic parameters are mostly constrained by the external measurements. The full results are given in Table~\ref{table:hadronic_parameters_fit}.  The accompanying correlation matrices can be found in Appendix~\ref{app:interpret_correlation_matrices}.

\subsection{Interpretation in terms of mixing parameters}

\begin{figure}[htb]
\centering
\includegraphics[width=0.495\textwidth]{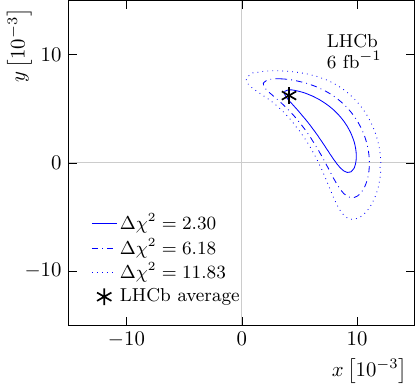}
\caption{\label{fig:xy}Likelihood contours of mixing parameters, assuming that $\CP$ violation in charm mixing can be neglected. The $\Delta\chi^2=2.38,~6.18,~11.83$ contours are indicated, corresponding to the $68.7,~95.0,~99.7\%$ confidence intervals for a Gaussian likelihood. The average of previous LHCb measurements for the mixing parameters is also shown, where the marker size is significantly larger than the current uncertainties~\cite{LHCb-CONF-2024-004}.}
\end{figure}

The likelihood scan in the $(x,y)$ plane is shown in Fig.~\ref{fig:xy} for the binned observables from the \CP-conserving analysis, including the constraints $\chi^2_\text{CLEO}$, $\chi^2_\text{BESIII}$ and $\chi^2_\text{R1}$. 
The combination $(x^2+y^2)$ is significantly more constrained than either of the mixing parameters individually, owing to the relatively large uncertainties on the phase differences in the bins. 
The individual mixing parameters are reported in Table~\ref{tab:mixingparameters} and are compatible with the averages of previous LHCb measurements within the $95\%$ confidence level~\cite{LHCb-CONF-2024-004}.

\begin{figure}[htb]
\includegraphics[width=0.495\textwidth]{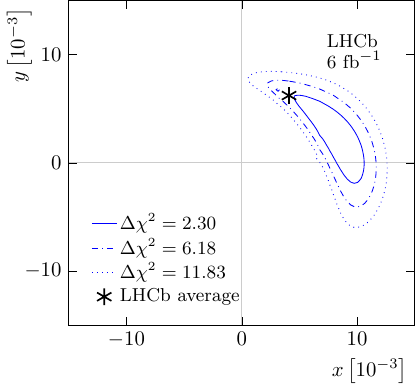}
\includegraphics[width=0.495\textwidth]{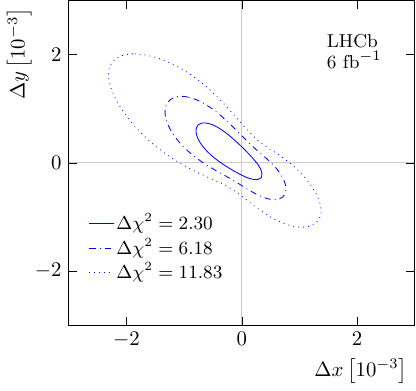}
\caption{\label{fig:cpvxy} Likelihood contours of (left)~mixing parameters, in the case that $\CP$ violation is allowed, and (right)~the $\CP$-violating parameters themselves. The $\Delta\chi^2=2.38,~6.18,~11.83$ contours are indicated, corresponding to the $68.7,~95.0,~99.7\%$ confidence intervals for a Gaussian likelihood. The average of previous LHCb measurements for the mixing parameters is also shown, where the marker size is significantly larger than the current uncertainties~\cite{LHCb-CONF-2024-004}.}
\end{figure}

A fit is also performed to the \CP-violating observables, with the same external constraints. The compatibility between measurements is found to be good, with the $\chi^2$ of the mixing fit found to be 5.6 for 10 degrees of freedom.
The results are presented in Table~\ref{tab:mixingparameters}.  Again, the results for $x$ and $y$ are compatible with the LHCb-average values~\cite{LHCb-CONF-2024-004}.   The results for $\Delta x$ and $\Delta y$ are compatible with zero and the hypothesis of \CP conservation.  The likelihood contours for both $(x,y)$ and $(\Delta x, \Delta y)$ are shown in Fig.~\ref{fig:cpvxy}.  The correlation matrices accompanying the results in Table~\ref{tab:mixingparameters} are given in Appendix~\ref{app:interpret_correlation_matrices}.

\begin{table}[htb]
\centering
\caption{Fitted results for the mixing parameters.
The measurements of the parameters $(x,y)$ are highly anticorrelated, with correlation coefficients of $\rho=-0.896,-0.847$ for $\CP$-conserving and $\CP$-violating fits, respectively. 
}
\def\arraystretch{1.2}

\begin{tabular}{lcccc} \toprule
& $x$ [\%]  & $y$ [\%]  &  $\phantom{-}\Delta x$ [\%] &  $\Delta y$ [\%] \\ \midrule
\CP-conserving fit  & \xmeascp   &  \ymeascp   &  ---  &  --- \\
\CP-violating fit   & \xmeas  & \ymeas  &  \dxmeas & \dymeas  \\\bottomrule
\end{tabular}
\label{tab:mixingparameters}
\end{table}

\section{Summary}
\label{sec:summary}

A study of charm mixing and \CP violation in \bothDz decays is performed with data collected by LHCb in $pp$ collisions, corresponding to an integrated luminosity of 6\invfb.   Observables characterising charm mixing are measured, both inclusive and in four bins of phase space, and are reported in Table~\ref{tab:mixingresults}. 

Taking the average values of the mixing parameters $x$ and $y$ from previous LHCb measurements as external inputs~\cite{LHCb-CONF-2024-004}, the inclusive measurement is interpreted in terms of $r$ and $\delta$, the magnitude and phase difference, respectively, of the ratio of the DCS to the CF amplitudes, averaged over phase space, and the coherence factor $\kappa$.  In combination with measurements performed at charm threshold~\cite{Evans2020_new,BESIII2021_new}, and with an earlier measurement from LHCb~\cite{LHCb-PAPER-2015-057}, this analysis yields
\begin{equation*}
\begin{split}
r &=  \left(\rgmeas\right) \times 10^{-2}  ,\\
\kappa &= \kgmeas  ,\\
\delta &= \left(\dgmeas\right)^\circ ,
\end{split}
\end{equation*}
which is a significant improvement in precision compared to the previous knowledge of these hadronic  parameters~\cite{BESIII2021_new}.   A similar conclusion is reached when these parameters are determined in the individual bins of phase space.   This improvement will be valuable in future measurements of the \CP-violating phase $\gamma$ in $B^- \to DK^-$, $D \to K^\pm \pi^\mp \pip \pim$  decays, where knowledge of the hadronic parameters is required.

A second interpretation is performed in which the hadronic parameters from previous measurements are taken as input, and the charm-mixing parameters are determined. The fit to the observables of Table~\ref{tab:mixingresults_cpv}, which derive from an analysis that allows for \CP violation, returns
\begin{equation*}
\begin{split}
x &=        \left( \phantom{-}\xmeas \right)\% , \\
y &=        \left( \phantom{-}\ymeas\right)\% , \\
\Delta x &= \left( \dxmeas          \right)\% ,\\
\Delta y &= \left( \phantom{-} \dymeas\right)\%.\\
\end{split}
\end{equation*}
These results are compatible with previous measurements~\cite{HFLAV23} and with the hypothesis of \CP conservation. The determination of the \CP violating component of the mixing parameters is one of the most precise to date.

The interpretation in terms of charm-mixing parameters is currently limited by the knowledge of the hadronic parameters of  \bothDz decays that come from external measurements, most importantly at charm threshold.  These measurements, and therefore the sensitivities of the mixing parameters, are expected to improve, as the BESIII experiment has now collected a data set substantially larger than that analysed in Ref.~\cite{BESIII2021_new}.  Moreover, the upgraded LHCb detector~\cite{LHCb-DP-2022-002} is accumulating larger samples of $pp$-collision data, which will allow this and similar studies to be repeated with greater precision.

\FloatBarrier

%% file: bin_definition.tex
\begin{table}  \caption{\label{bin_definitions} 
Measured values of $\kappa_{(i)}$ and $\delta_{(i)}$ inclusive over phase space (in phase-space bin $i$) from a combination of charm-threshold data and measurements performed by the LHCb collaboration in Run~1 of the LHC~\cite{Evans2020_new,BESIII2021_new,LHCb-PAPER-2015-057}, where the parameters are obtained by minimising a $\chi^2$ analogous to that discussed in Sect.~\ref{sec:interpretation}. The definition of the phase-space bins is given in Ref.~\cite{Evans2020_new}.}

  \centering
    \def\arraystretch{1.2}
\begin{tabular}{lrr}
\toprule
  Phase-space bin 
    & \multicolumn{1}{c}{$\kappa_{(i)}$}
    & \multicolumn{1}{c}{$\delta_{(i)} [^\circ]$}\\
\midrule
Inclusive &  $0.43 \aerr{0.07}{0.06}$ & $162\aerr{20}{18}$ \\
  1 &  $0.56 \aerr{0.19}{0.20}$ & $127\aerr{31}{14}$ \\
  2 &  $0.89 \aerr{0.11}{0.22}$ & $134\aerr{20}{12}$ \\
  3 &  $0.71 \aerr{0.11}{0.10}$ & $167\aerr{24}{19}$ \\
  4 &  $0.41 \aerr{0.25}{0.25}$ & $292\aerr{42}{18}$ \\
\bottomrule
\label{tab:coherence_factors}
\end{tabular}
\end{table}

%% file: acknowledgements.tex
\section*{Acknowledgements}
%
%
\noindent We express our gratitude to our colleagues in the CERN
accelerator departments for the excellent performance of the LHC. We
thank the technical and administrative staff at the LHCb
institutes.
We acknowledge support from CERN and from the national agencies:
ARC (Australia);
CAPES, CNPq, FAPERJ and FINEP (Brazil); 
MOST and NSFC (China); 
CNRS/IN2P3 (France); 
BMFTR, DFG and MPG (Germany);
INFN (Italy); 
NWO (Netherlands); 
MNiSW and NCN (Poland); 
MCID/IFA (Romania); 
MICIU and AEI (Spain);
SNSF and SER (Switzerland); 
NASU (Ukraine); 
STFC (United Kingdom); 
DOE NP and NSF (USA).
We acknowledge the computing resources that are provided by ARDC (Australia), 
CBPF (Brazil),
CERN, 
IHEP and LZU (China),
IN2P3 (France), 
KIT and DESY (Germany), 
INFN (Italy), 
SURF (Netherlands),
Polish WLCG (Poland),
IFIN-HH (Romania), 
PIC (Spain), CSCS (Switzerland), 
and GridPP (United Kingdom).
We are indebted to the communities behind the multiple open-source
software packages on which we depend.
Individual groups or members have received support from
Key Research Program of Frontier Sciences of CAS, CAS PIFI, CAS CCEPP, 
Fundamental Research Funds for the Central Universities,  and Sci.\ \& Tech.\ Program of Guangzhou (China);
Minciencias (Colombia);
EPLANET, Marie Sk\l{}odowska-Curie Actions, ERC and NextGenerationEU (European Union);
A*MIDEX, ANR, IPhU and Labex P2IO, and R\'{e}gion Auvergne-Rh\^{o}ne-Alpes (France);
Alexander-von-Humboldt Foundation (Germany);
ICSC (Italy); 
Severo Ochoa and Mar\'ia de Maeztu Units of Excellence, GVA, XuntaGal, GENCAT, InTalent-Inditex and Prog.~Atracci\'on Talento CM (Spain);
SRC (Sweden);
the Leverhulme Trust, the Royal Society and UKRI (United Kingdom).

%% file: appendix.tex

\section*{Appendices}

\appendix

\section{Statistical and systematic correlation matrices }
\label{app:stat_syst_cov}


The statistical correlation matrices for the \CP-conserving, phase-space inclusive and \CP-conserving, phase-space-binned mixing fits are given in Table~\ref{tab:fit_cor_stat}.
The systematic correlation matrix for these fits is given in Table~\ref{tab:cor_cpint}. 

The statistical correlation matrix for the \CP-violating phase-space-binned fit is given in Table~\ref{tab:cor_cpv_stat}.
The systematic correlation matrix for this fit is given in Table~\ref{tab:cor_cpv_syst}. 

\begin{table}[htb]
\centering
\caption[Statistical fit correlations]{
        Correlation coefficients of the statistical uncertainties for the \CP-conserving (top)~phase-space-inclusive fit and (bottom) phase-space-binned fit.
        }
\label{tab:fit_cor_stat}
\renewcommand{\arraystretch}{1.2}
\begin{tabular} { c c c c } 
\toprule
Parameter & $(x^2 + y^2)$  & $r$   & $\kappa y'$  \\ 
\midrule 
$(x^2 + y^2)$     & 1.000 &   &  \\
$r$   & 0.865  & 1.000   &  \\ 
$\kappa y'$           & 0.965  & 0.952    & 1.000 \\ 
\bottomrule 
\end{tabular} \\
\vskip 0.5cm
\begin{tabular} { c c c c c c c c c c c } 
\toprule
Parameter & $(x^2 + y^2)$  & $r_1$ & $r_2$ & $r_3$ & $r_4$ & $(\kappa y')_1$ & $(\kappa y')_2$ & $(\kappa y')_3$ & $(\kappa y')_4$ \\ 
\midrule 
$(x^2+y^2)$ & $1.000$ &  &  &  &  &  &  &  &  \\
$r_1$ & $0.646$ & $1.000$ &  &  &  &  &  &  &  \\
$r_2$ & $0.602$ & $0.399$ & $1.000$ &  &  &  &  &  &  \\
$r_3$ & $0.623$ & $0.411$ & $0.384$ & $1.000$ &  &  &  &  &  \\
$r_4$ & $0.712$ & $0.468$ & $0.436$ & $0.450$ & $1.000$ &  &  &  &  \\
$(\kappa y')_1$ & $0.872$ & $0.898$ & $0.536$ & $0.553$ & $0.628$ & $1.000$ &  &  &  \\
$(\kappa y')_2$ & $0.838$ & $0.553$ & $0.901$ & $0.532$ & $0.604$ & $0.743$ & $1.000$ &  &  \\
$(\kappa y')_3$ & $0.859$ & $0.567$ & $0.529$ & $0.892$ & $0.620$ & $0.762$ & $0.733$ & $1.000$ &  \\
$(\kappa y')_4$ & $0.910$ & $0.599$ & $0.559$ & $0.575$ & $0.902$ & $0.804$ & $0.774$ & $0.793$ & $1.000$ \\
\bottomrule 
\end{tabular} \\
\end{table}

\begin{table}[htb]
\centering
\caption[Systematic uncertainty correlations]{
        Correlation coefficients of the systematic uncertainties for the \CP-conserving (top)~phase-space-inclusive fit and (bottom) phase-space-binned fit.
        }
\label{tab:cor_cpint}
\renewcommand{\arraystretch}{1.2}
\begin{tabular} { c c c c } 
\toprule
Parameter & $(x^2 + y^2)$  & $r$   & $\kappa y'$  \\ 
\midrule 
$(x^2 + y^2)$     & 1.000 &   &  \\
$r$             & 0.875  & 1.000   &  \\ 
$\kappa y'$     & 0.955  & 0.969   & 1.000 \\ 
\bottomrule 
\end{tabular} \\
\vskip 0.5cm
\begin{tabular} { c c c c c c c c c c c } 
\toprule
Parameter & $(x^2 + y^2)$  & $r_1$ & $r_2$ & $r_3$ & $r_4$ & $(\kappa y')_1$ & $(\kappa y')_2$ & $(\kappa y')_3$ & $(\kappa y')_4$ \\ 
\midrule 
$(x^2+y^2)$ & $1.000$ &  &  &  &  &  &  &  &  \\
$r_1$ & $0.632$ & $1.000$ &  &  &  &  &  &  &  \\
$r_2$ & $0.661$ & $0.489$ & $1.000$ &  &  &  &  &  &  \\
$r_3$ & $0.541$ & $0.344$ & $0.374$ & $1.000$ &  &  &  &  &  \\
$r_4$ & $0.729$ & $0.454$ & $0.455$ & $0.420$ & $1.000$ &  &  &  &  \\
$(\kappa y')_1$ & $0.859$ & $0.905$ & $0.611$ & $0.456$ & $0.614$ & $1.000$ &  &  &  \\
$(\kappa y')_2$ & $0.881$ & $0.593$ & $0.910$ & $0.475$ & $0.627$ & $0.782$ & $1.000$ &  &  \\
$(\kappa y')_3$ & $0.837$ & $0.528$ & $0.556$ & $0.865$ & $0.626$ & $0.717$ & $0.736$ & $1.000$ &  \\
$(\kappa y')_4$ & $0.928$ & $0.582$ & $0.591$ & $0.514$ & $0.905$ & $0.793$ & $0.807$ & $0.788$ & $1.000$ \\
\bottomrule 
\end{tabular} \\
\end{table}

\newpage

\begin{sidewaystable}[t]
\caption[Correlation matrix of statistical uncertainties for CPV fit]{
        Correlation matrix of the statistical uncertainties for the \CP-violating phase-space-binned fit.
        }
\label{tab:cor_cpv_stat}
    \centering
    \scriptsize
    \setlength{\tabcolsep}{4.5pt}
        \renewcommand{\arraystretch}{1.2}
    \begin{tabular}{c r rrrr rrrr r rrrr rrrr}
    \toprule
& \multicolumn{1}{c}{$({x^+}^2\!+{y^+}^2$)}  & \multicolumn{1}{c}{$r_1^+$}  & \multicolumn{1}{c}{$r_2^+$}  & \multicolumn{1}{c}{$r_3^+$}  & \multicolumn{1}{c}{$r_4^+$}  & \multicolumn{1}{c}{$(\kappa y^{'+})_1$}  & \multicolumn{1}{c}{$(\kappa y^{'+})_2$}  & \multicolumn{1}{c}{$(\kappa y^{'+})_3$}  & \multicolumn{1}{c}{$(\kappa y^{'+})_4$}  & \multicolumn{1}{c}{$({x^-}^2\!+{y^-}^2)$}  & \multicolumn{1}{c}{$r_1^-$}  & \multicolumn{1}{c}{$r_2^-$}  & \multicolumn{1}{c}{$r_3^-$}  & \multicolumn{1}{c}{$r_4^-$}  & \multicolumn{1}{c}{$(\kappa y^{'-})_1$}  & \multicolumn{1}{c}{$(\kappa y^{'-})_2$}  & \multicolumn{1}{c}{$(\kappa y^{'-})_3$}  & \multicolumn{1}{c}{$(\kappa y^{'-})_4$}  \\
\midrule
$({x^+}^2\!+{y^+}^2)$ & $1.000$ &  &  &  &  &  &  &  &  &  &  &  &  &  &  &  &  &  \\
$r_1^+$ & $0.651$ & $1.000$ &  &  &  &  &  &  &  &  &  &  &  &  &  &  &  &  \\
$r_2^+$ & $0.608$ & $0.427$ & $1.000$ &  &  &  &  &  &  &  &  &  &  &  &  &  &  &  \\
$r_3^+$ & $0.628$ & $0.437$ & $0.411$ & $1.000$ &  &  &  &  &  &  &  &  &  &  &  &  &  &  \\
$r_4^+$ & $0.716$ & $0.490$ & $0.459$ & $0.472$ & $1.000$ &  &  &  &  &  &  &  &  &  &  &  &  &  \\
$(\kappa y'^+)_1$ & $0.873$ & $0.897$ & $0.553$ & $0.569$ & $0.645$ & $1.000$ &  &  &  &  &  &  &  &  &  &  &  &  \\
$(\kappa y'^+)_2$ & $0.847$ & $0.575$ & $0.891$ & $0.554$ & $0.627$ & $0.764$ & $1.000$ &  &  &  &  &  &  &  &  &  &  &  \\
$(\kappa y'^+)_3$ & $0.861$ & $0.583$ & $0.545$ & $0.891$ & $0.636$ & $0.775$ & $0.753$ & $1.000$ &  &  &  &  &  &  &  &  &  &  \\
$(\kappa y'^+)_4$ & $0.910$ & $0.611$ & $0.571$ & $0.588$ & $0.904$ & $0.815$ & $0.792$ & $0.804$ & $1.000$ &  &  &  &  &  &  &  &  &  \\
$({x^-}^2\!+{y^-}^2)$ & $-0.114$ & $-0.063$ & $-0.059$ & $-0.061$ & $-0.071$ & $-0.087$ & $-0.083$ & $-0.086$ & $-0.094$ & $1.000$ &  &  &  &  &  &  &  &  \\
$r_1^-$ & $-0.061$ & $-0.057$ & $-0.056$ & $-0.055$ & $-0.055$ & $-0.062$ & $-0.062$ & $-0.061$ & $-0.061$ & $0.663$ & $1.000$ &  &  &  &  &  &  &  \\
$r_2^-$ & $-0.056$ & $-0.056$ & $-0.055$ & $-0.054$ & $-0.053$ & $-0.059$ & $-0.059$ & $-0.058$ & $-0.057$ & $0.620$ & $0.424$ & $1.000$ &  &  &  &  &  &  \\
$r_3^-$ & $-0.060$ & $-0.055$ & $-0.054$ & $-0.054$ & $-0.053$ & $-0.059$ & $-0.059$ & $-0.059$ & $-0.059$ & $0.638$ & $0.436$ & $0.409$ & $1.000$ &  &  &  &  &  \\
$r_4^-$ & $-0.070$ & $-0.056$ & $-0.054$ & $-0.054$ & $-0.059$ & $-0.066$ & $-0.066$ & $-0.066$ & $-0.068$ & $0.731$ & $0.489$ & $0.457$ & $0.471$ & $1.000$ &  &  &  &  \\
$(\kappa y'^-)_1$ & $-0.085$ & $-0.062$ & $-0.059$ & $-0.060$ & $-0.066$ & $-0.080$ & $-0.079$ & $-0.079$ & $-0.082$ & $0.889$ & $0.898$ & $0.551$ & $0.568$ & $0.644$ & $1.000$ &  &  &  \\
$(\kappa y'^-)_2$ & $-0.083$ & $-0.062$ & $-0.059$ & $-0.059$ & $-0.065$ & $-0.079$ & $-0.079$ & $-0.078$ & $-0.080$ & $0.859$ & $0.572$ & $0.895$ & $0.551$ & $0.624$ & $0.760$ & $1.000$ &  &  \\
$(\kappa y'^-)_3$ & $-0.085$ & $-0.061$ & $-0.058$ & $-0.059$ & $-0.066$ & $-0.079$ & $-0.079$ & $-0.079$ & $-0.081$ & $0.879$ & $0.583$ & $0.544$ & $0.890$ & $0.637$ & $0.775$ & $0.751$ & $1.000$ &  \\
$(\kappa y'^-)_4$ & $-0.091$ & $-0.061$ & $-0.058$ & $-0.059$ & $-0.068$ & $-0.082$ & $-0.081$ & $-0.081$ & $-0.086$ & $0.929$ & $0.611$ & $0.569$ & $0.588$ & $0.903$ & $0.815$ & $0.789$ & $0.805$ & $1.000$ \\

    \bottomrule
    \end{tabular}
\end{sidewaystable}


\begin{sidewaystable}[t]
\caption[Correlation matrix of systematic uncertainties for CPV fit]{
        Correlation matrix of the systematic uncertainties for the \CP-violating phase-space-binned fit.
        }
\label{tab:cor_cpv_syst}
    \centering
    \scriptsize
        \setlength{\tabcolsep}{4.5pt}
    \renewcommand{\arraystretch}{1.2}
    \begin{tabular}{c r rrrr rrrr r rrrr rrrr}
    \toprule
& \multicolumn{1}{c}{$({x^+}^2\!+{y^+}^2)$}  & \multicolumn{1}{c}{$r_1^+$}  & \multicolumn{1}{c}{$r_2^+$}  & \multicolumn{1}{c}{$r_3^+$}  & \multicolumn{1}{c}{$r_4^+$}  & \multicolumn{1}{c}{$(\kappa y^{'+})_1$}  & \multicolumn{1}{c}{$(\kappa y^{'+})_2$}  & \multicolumn{1}{c}{$(\kappa y^{'+})_3$}  & \multicolumn{1}{c}{$(\kappa y^{'+})_4$}  & \multicolumn{1}{c}{$({x^-}^2\!+{y^-}^2)$}  & \multicolumn{1}{c}{$r_1^-$}  & \multicolumn{1}{c}{$r_2^-$}  & \multicolumn{1}{c}{$r_3^-$}  & \multicolumn{1}{c}{$r_4^-$}  & \multicolumn{1}{c}{$(\kappa y^{'-})_1$}  & \multicolumn{1}{c}{$(\kappa y^{'-})_2$}  & \multicolumn{1}{c}{$(\kappa y^{'-})_3$}  & \multicolumn{1}{c}{$(\kappa y^{'-})_4$}  \\
    \midrule
$({x^+}^2\!+{y^+}^2)$ & $1.000$ &  &  &  &  &  &  &  &  &  &  &  &  &  &  &  &  &  \\
$r_1^+$ & $0.620$ & $1.000$ &  &  &  &  &  &  &  &  &  &  &  &  &  &  &  &  \\
$r_2^+$ & $0.653$ & $0.422$ & $1.000$ &  &  &  &  &  &  &  &  &  &  &  &  &  &  &  \\
$r_3^+$ & $0.582$ & $0.367$ & $0.404$ & $1.000$ &  &  &  &  &  &  &  &  &  &  &  &  &  &  \\
$r_4^+$ & $0.689$ & $0.436$ & $0.447$ & $0.449$ & $1.000$ &  &  &  &  &  &  &  &  &  &  &  &  &  \\
$(\kappa y'^+)_1$ & $0.862$ & $0.893$ & $0.565$ & $0.496$ & $0.603$ & $1.000$ &  &  &  &  &  &  &  &  &  &  &  &  \\
$(\kappa y'^+)_2$ & $0.883$ & $0.548$ & $0.902$ & $0.520$ & $0.608$ & $0.766$ & $1.000$ &  &  &  &  &  &  &  &  &  &  &  \\
$(\kappa y'^+)_3$ & $0.851$ & $0.517$ & $0.561$ & $0.868$ & $0.612$ & $0.725$ & $0.755$ & $1.000$ &  &  &  &  &  &  &  &  &  &  \\
$(\kappa y'^+)_4$ & $0.919$ & $0.564$ & $0.591$ & $0.553$ & $0.890$ & $0.792$ & $0.810$ & $0.795$ & $1.000$ &  &  &  &  &  &  &  &  &  \\
$({x^-}^2\!+{y^-}^2)$ & $1.000$ & $0.620$ & $0.653$ & $0.582$ & $0.689$ & $0.862$ & $0.883$ & $0.851$ & $0.919$ & $1.000$ &  &  &  &  &  &  &  &  \\
$r_1^-$ & $0.620$ & $1.000$ & $0.422$ & $0.367$ & $0.436$ & $0.894$ & $0.548$ & $0.517$ & $0.564$ & $0.620$ & $1.000$ &  &  &  &  &  &  &  \\
$r_2^-$ & $0.654$ & $0.422$ & $1.000$ & $0.404$ & $0.447$ & $0.565$ & $0.903$ & $0.561$ & $0.592$ & $0.654$ & $0.422$ & $1.000$ &  &  &  &  &  &  \\
$r_3^-$ & $0.582$ & $0.367$ & $0.405$ & $1.000$ & $0.449$ & $0.496$ & $0.519$ & $0.868$ & $0.553$ & $0.582$ & $0.367$ & $0.404$ & $1.000$ &  &  &  &  &  \\
$r_4^-$ & $0.689$ & $0.436$ & $0.447$ & $0.449$ & $1.000$ & $0.603$ & $0.608$ & $0.612$ & $0.890$ & $0.689$ & $0.436$ & $0.447$ & $0.449$ & $1.000$ &  &  &  &  \\
$(\kappa y'^-)_1$ & $0.862$ & $0.893$ & $0.565$ & $0.496$ & $0.603$ & $1.000$ & $0.766$ & $0.725$ & $0.792$ & $0.862$ & $0.893$ & $0.565$ & $0.496$ & $0.603$ & $1.000$ &  &  &  \\
$(\kappa y'^-)_2$ & $0.883$ & $0.548$ & $0.902$ & $0.519$ & $0.607$ & $0.766$ & $1.000$ & $0.755$ & $0.810$ & $0.883$ & $0.548$ & $0.902$ & $0.519$ & $0.607$ & $0.766$ & $1.000$ &  &  \\
$(\kappa y'^-)_3$ & $0.851$ & $0.517$ & $0.561$ & $0.868$ & $0.612$ & $0.725$ & $0.755$ & $1.000$ & $0.795$ & $0.851$ & $0.517$ & $0.562$ & $0.868$ & $0.612$ & $0.725$ & $0.755$ & $1.000$ &  \\
$(\kappa y'^-)_4$ & $0.919$ & $0.564$ & $0.591$ & $0.553$ & $0.890$ & $0.792$ & $0.810$ & $0.795$ & $1.000$ & $0.919$ & $0.564$ & $0.592$ & $0.553$ & $0.890$ & $0.792$ & $0.810$ & $0.795$ & $1.000$ \\
    \bottomrule
    \end{tabular}
\end{sidewaystable}

\FloatBarrier 
\section{Correlation matrices of interpretation fits}
\label{app:interpret_correlation_matrices}

The correlation matrices from the fit used to determine the hadronic parameters $(r,\kappa,\delta)$ are given in Table~\ref{tab:cor_coherence_factors_global} and \ref{tab:cor_coherence_factors} for phase-space inclusive and binned parameters, respectively. 
The correlation matrix for the fit to determine the mixing parameters in the $\CP$-violating fit is given in Table~\ref{tab:xy_cpv_correlation}, while the only nontrivial component of the $\CP$-conserving correlation matrix is the correlation between $x$ and $y$, which is $-0.896$. 

\begin{table}[htb]
\centering
\caption{\label{tab:cor_coherence_factors_global}Correlation matrix for the phase-space inclusive interpretation fit, where the charm-mixing parameters are constrained to the global averages and thus used for the determination of the hadronic parameters.}
    \renewcommand{\arraystretch}{1.2}

\begin{tabular}{c r r r}

\toprule
& \multicolumn{1}{c}{$r$} & \multicolumn{1}{c}{$\delta$} & \multicolumn{1}{c}{$\kappa$} \\
\midrule
$r$ & $1.000$ &  &  \\
$\delta$ & $0.340$ & $1.000$ &  \\
$\kappa$ & $-0.505$ & $0.345$ & $1.000$ \\
\bottomrule
\end{tabular}
\end{table}

\begin{table}[htb]
\caption{\label{tab:cor_coherence_factors}Correlation matrix for the binned interpretation fit, where the charm-mixing parameters are constrained to the global averages and thus used for the determination of the binned hadronic parameters.}
\scriptsize
\setlength{\tabcolsep}{5.4pt}
\centering
\renewcommand{\arraystretch}{1.2}
\begin{tabular}{c rrrr rrrr rrrr }
\toprule
  &  \multicolumn{1}{c}{$r_1$} 
   & \multicolumn{1}{c}{$r_2$} 
   & \multicolumn{1}{c}{$r_3$} 
   & \multicolumn{1}{c}{$r_4$} 
   & \multicolumn{1}{c}{$\delta_1$} 
   & \multicolumn{1}{c}{$\delta_2$} 
   & \multicolumn{1}{c}{$\delta_3$} 
   & \multicolumn{1}{c}{$\delta_4$} 
   & \multicolumn{1}{c}{$\kappa_1$} 
   & \multicolumn{1}{c}{$\kappa_2$} 
   & \multicolumn{1}{c}{$\kappa_3$} 
   & \multicolumn{1}{c}{$\kappa_4$}  \\
  \midrule
$r_1$ & $1.000$ &  &  &  &  &  &  &  &  &  &  &  \\
$r_2$ & $0.006$ & $1.000$ &  &  &  &  &  &  &  &  &  &  \\
$r_3$ & $0.022$ & $0.012$ & $1.000$ &  &  &  &  &  &  &  &  &  \\
$r_4$ & $0.002$ & $-0.002$ & $0.031$ & $1.000$ &  &  &  &  &  &  &  &  \\
$\delta_1$ & $0.186$ & $-0.024$ & $0.097$ & $-0.016$ & $1.000$ &  &  &  &  &  &  &  \\
$\delta_2$ & $-0.033$ & $0.382$ & $0.057$ & $0.018$ & $0.027$ & $1.000$ &  &  &  &  &  &  \\
$\delta_3$ & $0.046$ & $0.024$ & $0.626$ & $0.032$ & $0.213$ & $0.126$ & $1.000$ &  &  &  &  &  \\
$\delta_4$ & $0.037$ & $0.032$ & $0.002$ & $-0.244$ & $0.022$ & $0.011$ & $-0.007$ & $1.000$ &  &  &  &  \\
$\kappa_1$ & $-0.582$ & $-0.002$ & $-0.143$ & $-0.021$ & $-0.716$ & $-0.130$ & $-0.256$ & $-0.034$ & $1.000$ &  &  &  \\
$\kappa_2$ & $-0.028$ & $-0.730$ & $-0.105$ & $-0.066$ & $-0.092$ & $-0.417$ & $-0.193$ & $-0.025$ & $0.337$ & $1.000$ &  &  \\
$\kappa_3$ & $0.031$ & $0.007$ & $0.101$ & $-0.002$ & $0.175$ & $0.030$ & $0.733$ & $-0.011$ & $-0.158$ & $-0.061$ & $1.000$ &  \\
$\kappa_4$ & $-0.008$ & $-0.012$ & $0.041$ & $0.805$ & $0.008$ & $0.056$ & $0.075$ & $-0.505$ & $-0.096$ & $-0.132$ & $0.015$ & $1.000$ \\
\bottomrule
\end{tabular}
\end{table}

\begin{table}[htb]
\centering 
    \renewcommand{\arraystretch}{1.2}

\caption{\label{tab:xy_cpv_correlation}Correlation matrix of the charm-mixing parameters for the fits allowing for $\CP$ violation.}
\begin{tabular}{c rrrr} 
\toprule
& \multicolumn{1}{c}{$x$} & \multicolumn{1}{c}{$y$} & \multicolumn{1}{c}{$\Delta x$} & \multicolumn{1}{c}{$\Delta y$}  \\
\midrule
$x$ & $1.000$ &  &  &  \\
$y$ & $-0.847$ & $1.000$ &  &  \\
$\Delta x$ & $-0.071$ & $0.061$ & $1.000$ &  \\
$\Delta y$ & $0.241$ & $-0.281$ & $-0.799$ & $1.000$ \\
\bottomrule
\end{tabular}
\end{table}

%% file: Authorship_LHCb-PAPER-2025-029.tex
\centerline
{\large\bf LHCb collaboration}
\begin
{flushleft}
\small
R.~Aaij$^{38}$\lhcborcid{0000-0003-0533-1952},
A.S.W.~Abdelmotteleb$^{57}$\lhcborcid{0000-0001-7905-0542},
C.~Abellan~Beteta$^{51}$\lhcborcid{0009-0009-0869-6798},
F.~Abudin{\'e}n$^{57}$\lhcborcid{0000-0002-6737-3528},
T.~Ackernley$^{61}$\lhcborcid{0000-0002-5951-3498},
A. A. ~Adefisoye$^{69}$\lhcborcid{0000-0003-2448-1550},
B.~Adeva$^{47}$\lhcborcid{0000-0001-9756-3712},
M.~Adinolfi$^{55}$\lhcborcid{0000-0002-1326-1264},
P.~Adlarson$^{85}$\lhcborcid{0000-0001-6280-3851},
C.~Agapopoulou$^{14}$\lhcborcid{0000-0002-2368-0147}
% ...

{\footnotesize \it

% ...
$^{14}$Universit{\'e} Paris-Saclay, CNRS/IN2P3, IJCLab, Orsay, France\\
% ...
$^{38}$Nikhef National Institute for Subatomic Physics, Amsterdam, Netherlands\\
% ...
$^{47}$Instituto Galego de F{\'\i}sica de Altas Enerx{\'\i}as (IGFAE), Universidade de Santiago de Compostela, Santiago de Compostela, Spain\\
% ...
$^{51}$Physik-Institut, Universit{\"a}t Z{\"u}rich, Z{\"u}rich, Switzerland\\
% ...
$^{55}$H.H. Wills Physics Laboratory, University of Bristol, Bristol, United Kingdom\\
% ...
$^{57}$Department of Physics, University of Warwick, Coventry, United Kingdom\\
% ...
$^{60}$School of Physics and Astronomy, University of Glasgow, Glasgow, United Kingdom\\
$^{61}$Oliver Lodge Laboratory, University of Liverpool, Liverpool, United Kingdom\\
% ...
$^{69}$Syracuse University, Syracuse, NY, United States\\
% ...
$^{85}$Department of Physics and Astronomy, Uppsala University, Uppsala, Sweden, associated to $^{60}$\\
% ...
}
\end{flushleft}